\documentclass[12pt,preprint]{aastex}
\begin{document}
\tighten
\title{Laboratory Experiments, Numerical Simulations, and Astronomical 
Observations of Deflected Supersonic Jets: Application to HH 110}

\author{
	P. Hartigan \altaffilmark{1}, 
	J. M. Foster \altaffilmark{2}
	B. H. Wilde \altaffilmark{3}
	R. F. Coker \altaffilmark{3}
	P. A. Rosen \altaffilmark{2}
	J. F. Hansen \altaffilmark{4}
	B. E. Blue \altaffilmark{5}
	R. J. R. Williams \altaffilmark{2}
	R. Carver \altaffilmark{1}
	A. Frank \altaffilmark{6}
	}

\vspace{1.0cm}

\altaffiltext{1}{Rice University, Department of Physics and Astronomy,
6100 S. Main, Houston, TX 77521-1892} 

\altaffiltext{2}{AWE, Aldermaston, Reading Berkshire, RG7 4PR, UK}

\altaffiltext{3}{Los Alamos National Laboratory, Los Alamos, NM 87545}

\altaffiltext{4}{Lawrence Livermore National Laboratory, 7000 East Avenue,
Livermore, CA 94550}

\altaffiltext{5}{General Atomics, 3550 General Atomics Court, San Diego, CA 92121-1122}

\altaffiltext{6}{University of Rochester, Department of Physics and Astronomy,
Rochester, NY 14627-0171} 

\begin{abstract}

Collimated supersonic flows in laboratory experiments
behave in a similar manner to astrophysical jets provided that radiation,
viscosity, and thermal conductivity are unimportant in the laboratory
jets, and that the experimental and astrophysical jets share similar dimensionless
parameters such as the Mach number and the ratio of the density between the jet and
the ambient medium.  When these conditions apply, laboratory jets
provide a means to study their astrophysical counterparts for a variety
of initial conditions, arbitrary viewing angles, and different times,
attributes especially helpful for interpreting astronomical images
where the viewing angle and initial conditions are fixed and the time domain is limited. 
Experiments are also a powerful way to test numerical
fluid codes in a parameter range where the codes must perform well.
In this paper we combine images from a series of laboratory experiments of deflected
supersonic jets with numerical simulations and new spectral observations of an
astrophysical example, the young stellar jet HH~110.
The experiments provide key insights into how deflected jets evolve in 3-D, particularly within 
working surfaces where multiple subsonic shells and filaments form, and along
the interface where shocked jet material penetrates into and destroys the obstacle
along its path. The experiments also underscore the importance of the 
viewing angle in determining what an observer will see.  
The simulations match the experiments so well that we can use the simulated velocity maps
to compare the dynamics in the experiment with those implied by the astronomical spectra. 
The experiments support a model where the observed shock structures in HH 110
form as a result of a pulsed driving source rather than from weak shocks that may arise in
the supersonic shear layer between the Mach disk and bow shock of the jet's working surface.

\keywords{Hydrodynamics, Methods: laboratory, ISM: jets and outflows, Herbig-Haro objects}

\end{abstract}

\section{Introduction}

Collimated supersonic jets originate from a variety of astronomical sources,
including active galactic nuclei \citep{agnref}, several kinds of interacting
binaries \citep{binaryjetref}, young stars \citep{ysojetref}, and even planetary
nebulae \citep{pnjetref}. 
Most current jet research focuses on how accretion disks accelerate and collimate
jets, or on understanding the dynamics of the jet as it generates shocks along its beam and in the surrounding medium.
Both areas of research have broad implications for astrophysics.
Models of accretion disks typically employ magnetized jets to remove the angular momentum
of the accreting material. The distribution and transport of the angular
momentum in an accretion disk affects its mass accretion rate, mixing, temperature profile,
and density structure, and in the case of young stars, also helps to define the characteristics
of the protoplanetary disk that remains after accretion ceases.  At larger distances from the source,
shock waves in jets clear material from the surrounding
medium, provide insights into the nature of density and velocity perturbations in the flow,
and enable dynamical studies of mixing, turbulence and shear.

Jets from young stars are particularly good testbeds for
investigating all aspects of the physics within collimated supersonic flows
\citep[see][for a review]{ray07}.
Shock velocities within stellar jets are low enough that the gas cools by
radiating emission lines rather than by expanding. Relative fluxes of the 
emission lines determine the density, temperature, and ionization of the postshock
gas, while the observed Doppler shifts and emission line profiles define the
radial velocities and nonthermal motions within the jet 
\citep[see][for a review]{hartigan08}. Moreover, many stellar
jets are located relatively close to the Earth, so that one can observe
proper motions in the plane of the sky from observations separated by several
years \citep{heathcote92}.
Combining this information with radial velocity measurements gives
the orientation of the flow to the line of sight.
Using the Hubble Space Telescope, one can observe morphological
changes of knots within jets, and follow how these changes evolve in real time 
\citep{hartigan01}.

Results from these studies show that internal shock waves, driven by velocity
variations in the flow, sweep material in jets into a series of dense knots. 
Typical internal shock velocities are $\sim$ 40 km$\,$s$^{-1}$, or $\sim$ 20\%\ of
the flow speed. In several cases it is easy to identify both the bow shock and
the Mach disk from emission line images. Because stellar jets are mostly neutral,
strong H$\alpha$ emission occurs at the shock front where neutral H is collisionally
excited \citep{heathcote96}.
Forbidden line radiation occurs in a spatially-extended cooling zone in
the postshock material. Temperatures immediately behind the internal shocks can
exceed $10^5$K, but the gas in the forbidden-line-emitting cooling zone
is typically 8000~K.

Young stars show a strong correlation between accretion and outflow,
leading to the idea that accretion disks power the outflows \citep{heg95,cabrit07}. 
Most current models use
magnetic fields in the disk to launch a fraction of the accreting material from
the disk into a collimated magnetized jet \citep{ferreira06}.
Stellar jets often precess,
and there is some evidence that they rotate \citep{ray07}, although rotation
signatures are difficult to measure because the rotational velocities are typically only
a few percent of the flow speeds and precession can mimic rotational signatures
\citep{cerquieriaref}.  In all cases proper motion measurements show that jets move
radially away from the source.  Jets show no dynamical evidence for kink instabilities,
and in fact while magnetic fields may dominate in the acceleration regions of jets,
they appear to play a minor role in the dynamics at the distances where most jet
knots are observed \citep{hartigan07}. In the cooling zones behind the shock waves
the plasma $\beta$ can drop below unity,
so that magnetic fields dominate thermal pressure in those areas. However,
the magnetic pressure is small compared with the ram pressure of the jet. 

While more unusual than internal working surfaces produced by velocity perturbations,
shocks also occur when jets collide with dense obstacles such as a molecular cloud.
When the obstacle is smaller than the jet radius, it becomes entrained by the jet, and
a reverse bow shock or `cloudlet shock' forms around the obstacle \citep{schwartz78,l1551}.
Alternatively, when a large obstacle like a molecular cloud deflects the jet, a quasi-stationary
deflection shock at the impact point forms, followed by a spray of shocked jet material
downstream. The classic example of such a jet is HH 110 \citep{riera03,lopez05}.

Though the observations summarized above provide a great deal of information
about stellar jets, several important questions remain unanswered. The basic 
mechanism by which disks load material onto field lines (assuming the MHD disk
scenario is correct), and the overall geometry of this wind is unknown, and the 
roles of reconnection and ambipolar diffusion in heating the jet close to the
star are unclear. At larger distances, the magnetic geometry and its 
importance in shaping the internal working surfaces is poorly-constrained, as
are the time scales and spatial scales associated with mixing in supersonic shear layers
and working surfaces. The inherently clumpy nature of jets also affects the flow
dynamics and observed properties of jets in uncertain ways, and the degree to
which fragmentation and turbulence influence the morphologies of jets is unknown.

Developing laboratory analogs of stellar jets could help significantly in
addressing the questions above. Observations of a specific astronomical jet
are restricted to a small range of times and to a particular observing angle,
while laboratory experiments have no such restrictions. In principle, one could
explore a wide range of initial collimations, velocity and density structures
within the jet, as well as densities, geometries, and magnetic field configurations
with laboratory experiments.  The experiments also provide a powerful and flexible way
to test 3-D numerical fluid codes, and to investigate how real flows develop complex
morphologies in 3-D.

The challenge is to design an experiment that is relevant to the astrophysical
case of interest. Laboratory experiments differ by 15 $-$ 20 orders of magnitude
in size, density and timescale from stellar jets, but because the
Euler equations that govern fluid dynamics involve only three variables,
time, density, and velocity, the behavior of the fluid is determined 
primarily by dimensionless numbers such as the Mach number (supersonic or subsonic),
Reynolds number (viscous or inertial), and Peclet number (importance of thermal conduction).
If the experiments behave as a fluid and have similar dimensionless fluid numbers
as those of stellar jets, then the experiments should scale well to the astrophysical case
\citep{ryutov99}.
Other parameters, such as magnetic fields and radiational cooling are more difficult
to match, and it is impossible to study the non-LTE excitation physics in the lab
because the critical density for collisional deexcitation is not scalable. 
In any case, the materials are markedly different between the experiments and stellar jets,
so it is not possible to study emission line ratios in any meaningful way
with current laboratory capabilities. Hence, at present the main utility of laboratory
experiments of jets is to clarify how complex supersonic flows evolve with time.

Laboratory work relevant to stellar jets is an emerging area of research, and several
papers have appeared recently which address various aspects of supersonic flows
in the appropriate regime \cite[see][for a review]{remington06}.
\citet{hansen07} observed how a strong planar shock
wave disrupts a spherical obstacle and tested numerical models of the process, and
\citet{loupias} generated a laboratory jet with a Mach number similar to that of a stellar jet.
In a different approach, \citet{lebedev04} and \citet{ampleford07}
used a conical array of wires at the Magpie facility to drive a magnetized jet, and 
explored the geometry of the deflection shock formed
as a jet impacts a crosswind. Laboratory experiments have also
recently studied the physics associated with
instabilities along supernova blast waves \citep{snrref,drake09}, and the dynamics
within supernovae explosions \citep{snrcoreref}. 

In this paper we present the results of a suite of experiments which deflect a
supersonic jet from a spherical obstacle, where the dimensionless fluid parameters are
similar to those present in stellar jets. In section 2, we describe our experimental
design, consider how the experiments scale to astrophysical jets,
demonstrate that the experiments
are reproducible, and report how the observed flows change as
one varies several parameters, including
the distance between the axis of the jet and the obstacle, the time delay,
the density probed with different backlighters, and the viewing angle.
The numerical work is summarized in section 3.
Detailed calculations with the 3-D RAGE code reproduce all of the major
observed morphologies well. In section 4, we present new high-resolution
optical spectra of the shocked wake of the HH 110 protostellar jet, and a new
wide-field H$_2$ image of the region.
These observations quantify how the internal dynamics of the gas behave as material
flows away from the deflection shock and show 
how the jet entrains material from the molecular cloud core.
Finally, in section 5 we consider how the experiments and the
simulations from RAGE provide new insights into the internal
dynamics of deflected supersonic jets, especially in the regions of the working surface
and at the interface where jet material entrains and accelerates the obstacle.

\section{Laser Experiments of Deflected Supersonic Jets} %2

\subsection{Experimental Design} %2.1

Figure 1 shows the experimental assembly we used to collide a supersonic jet into 
a spherical obstacle. The design consists of a
125 $\mu$m thickness titanium disk in direct contact with a 700 $\mu$m thickness
titanium washer with a central, 300 $\mu$m diameter hole. The surface of the
disk is heated by the thermal (soft x-ray) radiation from a hohlraum laser
target, which itself is heated by 12~beams of the Omega laser at the University of 
Rochester \citep{soures96}. X-ray driven
ablation of the surface of the disk creates a near-planar
shock within the disk and washer assembly, and the subsequent breakout of this
shock from the inner surface of the disk results in the directed outflow of a
plug of dense, shock-heated titanium plasma through the cylindrical hole
in the washer. This outflow is further collimated by the hole in the washer,
and directed into an adjacent block of low-density (0.1~g$\,$cm$^{-3}$) polymer foam
within which it propagates to a distance of $\sim$ 2~mm in $\sim$ 200~ns.

After the primary jet forms, the ablation-driven shock continues
to progress through the titanium target assembly and along the sides of the hole
in the washer. As the hole collapses inward a
secondary jet of material forms by a process analogous to that occurring in a
shaped-charge explosive. The secondary jet forms 
after the primary jet, and propagates into the
high-pressure cocoon of material already within the polymer foam.
At later times, the shock propagates from the surrounding titanium washer
into the foam, and causes the interface between the titanium
washer and the hydrocarbon foam to move.  A bow shock runs ahead of the
jets into the foam.  Both jets form as a result of
hydrodynamic phenomena alone; there is no significant magnetic field and
associated magneto-hydrodynamics, and the temperature of the jet is
sufficiently low for thermal-conduction and radiative energy losses to be
insignificant. Hence, the hydrodynamic phenomena which determine how the jet
forms and evolves are scalable (see Section 2.2) to jets of
very different dimensions that evolve over very different timescales. 

The experimental assembly described above
is identical in many respects to that used in experiments we have reported
previously \citep{foster05}, but with the following two significant
differences \citep[see][for further details]{rosen06, coker07}: (1)
in the present case we use indirect (hohlraum) drive, instead of the
direct laser drive used in our earlier work; and (2) the foam medium through
which the jet propagates contains an obstacle (a ball of CH).
The hohlraum drive enables us to obtain greater spatial
uniformity of the ablation of the titanium surface, and thereby generates a
jet of improved cylindrical symmetry. The (optional) addition of an obstacle
in the foam along the path of the jet medium makes it possible to set up,
and in a controlled manner test, how the flow behaves when the
two-dimensional cylindrical symmetry is broken.
A pinhole-apertured, laser-produced-plasma, x-ray backlighting
source projects an image of the jet onto radiographic film for later,
detailed analysis. 

Details of the experimental setup are as follows.
A 1600 $\mu$m diameter, 1200 $\mu$m length
(internal dimensions) cylindrical gold hohlraum target with a single
1200 $\mu$m diameter laser-entry hole \citep[see also][]{foster02}
generates the radiation drive.  The experimental package 
mounts over an 800 $\mu$m diameter hole in the end wall of the hohlraum,
immediately opposite the laser entry hole. This axisymmetric
configuration enables us to model the assembly at high resolution
using two-dimensional radiation hydrocodes during the stages
of formation and early-time evolution of the jet; the later
stages of three-dimensional hydrodynamics are modeled by
linking to a three-dimensional hydrocode, albeit at lower
spatial resolution. The hohlraum is heated by 12~beams of the Omega 
laser with a total energy of 6~kJ in a 1~ns duration, constant power
laser pulse of 0.35~$\mu$m wavelength. The resulting peak radiation
temperature in the hohlraum measured with a filtered
x-ray diode diagnostic (Dante) is 190 $-$ 200~eV \citep{foster02}. 

The titanium experimental assembly is made of an alloy with 90\% titanium,
6\% aluminium, and 4\% vanadium. The diamond-polished surfaces of the
hole and the planar surfaces of the components each have a 0.05 $-$ 0.3 $\mu$m
peak-valley, and 0.01 $-$ 0.03 $\mu$m RMS, surface finish. We place a 100~$\mu$m thickness,
500 $\mu$m\ diameter gold `cookie-cutter' disk between the hohlraum
and the titanium disk to control the area of the titanium disk illuminated by
x-rays, and to control the time it takes shocks to propagate into the titanium washer.
By this means we are able to adjust, to some extent, the relative
importance to the overall hydrodynamics of the primary and secondary jets,
and the late-time motion of the titanium/foam interface that surround both jets.
The medium through which the jets propagate is a 4 mm diameter, 6 mm length
cylinder of resorcinol-formaldehyde (C$_{15}$H$_{12}$O$_4$; hereafter RF) foam, of
0.1 g$\,$cm$^{-3}$ density.

We used RF foam because it has a very small ($<$ 1 $\mu$m) pore size.
The obstacle in the foam is a 1 mm diameter, solid polystyrene (1.03 g$\,$cm$^{-3}$ density) sphere,
supported by a small-diameter, silicon-carbide-coated tungsten stalk.
The axial position and radial offset (impact parameter) of the sphere
within the monomer material is set (approximately) before polymerisation
of the foam, but is determined accurately after polymerisation by 
inspecting each experimental assembly radiographically. Typically, the axial position
of the center of the ball, relative to the titanium-to-foam interface
is 800 $-$ 1000 $\mu$m, and the impact parameter (perpendicular distance from
the axis of the jet to the centre of the ball) is in the range 300 $-$ 500 $\mu$m.
We could determine these quantities with an accuracy of approximately
$\pm$ 10 $\mu$m, by analyzing pre-shot radiographs of the
experimental assembly. The process of target fabrication is to some
extent non-repeatable, and necessitates a separate hydrocode simulation of each
experimental shot once the target dimensions are known.
A sequence of experimental shots, diagnosed
at different times, thus measures the hydrodynamic behavior of
several very similar (but not strictly identical) target assemblies. 

A thin-foil, transition-metal, laser target illuminated
with 2 $-$ 5 beams of the Omega laser creates the
x-ray point backlighting source we use to diagnose the hydrodynamics.
Each beam provides 400~J of
energy in a 1~ns duration laser pulse, focused into a 600~$\mu$m diameter spot
(spot size determined by use of a random-phase plate).
X-ray emission from this laser-produced plasma passes through a laser-machined pinhole of
typically 10~$-$~20 $\mu$m diameter in 50~$\mu$m thickness tantalum foil, and generates
a backlighting source of size comparable to the pinhole aperture.
He-like resonance-line radiation of the backlighter target material
dominates the spectrum of this x-ray backlighting source, 
and the choice of backlighter targets of typically titanium, iron or zinc results in
He-like resonance line radiation of, respectively, 4.75, 6.7 and 9.0 keV.

Radiation from the point x-ray backlighting source creates a
point-projection (shadow) image of the experimental assembly on Kodak
DEF x-ray film, with approximately 10-times magnification.
Temporal resolution is determined by the duration of the x-ray
backlighting source (very nearly equal to the laser pulse length).
Motion blurring is insignificant for the 1~ns duration backlighting source.
The time delay between the laser
beams heating the hohlraum target, and the laser beams incident on the
x-ray backlighting target is varied from shot to shot to build up a
sequence of x-ray images of the jet hydrodynamics.

\subsection{Scaling the Experiments to Stellar Jets}%2.2

The size scales, times, densities and pressures
within the Omega experiments differ markedly from those
present in stellar jets.
For the laboratory results to be meaningful the fluid dynamical variables
must scale well, and the experiment should resemble the overall density 
and velocity structures present in astrophysical jets. We can connect
the fluid dynamics of the astrophysical and experimental cases through the
Euler equations for a polytropic gas \citep[e.g.][]{ll87}

\begin{equation}\label{eq:mass}
{{\partial \rho}\over{\partial t}} + \nabla \cdot \left(\rho\bf{v}\right)
= 0
\end{equation}
\begin{equation}\label{eq:momentum}
\rho\left({{\partial\bf{v}}\over{\partial t}} + \bf{v}\cdot\bf{\nabla v}
\right) + \nabla {\rm P} = 0
\end{equation}
\begin{equation}\label{eq:energy}
{{\partial {\rm P}}\over{\partial t}} + \gamma {\rm P} \nabla \cdot \bf{v} + \bf{v}
\cdot \nabla {\rm P} = 0,
\end{equation}

\noindent
where $\rho$ is the density, $\bf{v}$ is the velocity, and P is the gas pressure. 
\citet{ryutov99} showed that these equations are invariant to the rescaling
\begin{equation}
r^\prime = ar ;\ \ \  \rho^\prime = b\rho ;\ \ \  P^\prime = cP
\end{equation}
\noindent 
where a, b, and c are constants, provided one also rescales the time as
\begin{equation}
t^\prime = a\sqrt{{b\over c}}t.
\end{equation}
\noindent
With this scaling, the velocity transforms as
\begin{equation}
V^\prime = \sqrt{{c\over b}}V.
\end{equation}
\noindent
Solutions to the Euler equations will be identical in a dimensionless sense provided
equation 6 holds.

Hence, to verify that the experiment scales to the astrophysical case 
we need to estimate the pressures,
temperatures, and velocities in both the experiment and in the HH 110 system.
The parameters in the jet differ from those in the working surface where
the jet deflects from the obstacle, and from those in the working surface where the bow
shock impacts the ambient medium, so we must consider these regions separately.
The density and temperature vary throughout stellar jets as material encounters
weak shocks, heats, and cools, so we are interested in order of magnitude estimates of
these quantites for the scaling estimates.  As the discussion below will show,
the two working surfaces scale well in our experiment but the jet scales less well. 

Consider the parameters in the jet first.
Unlike most other jets from young stars, the HH~270 jet that collides with the molecular
cloud to produce HH~110 is rather ill-defined, and consists of  
several faint wisps that resemble weak bow shocks \citep{choi06}.
The electron density and ionization fraction in this part of the flow is poorly-constrained
by observations, but an electron density of $10^3$ cm$^{-3}$ is typical for faint
shocked structures of this kind. Taking a typical ionization fraction of 10\%
we obtain a total density of $10^4$ cm$^{-3}$, or about $2\times 10^{-20}$ g$\,$cm$^{-3}$.
These values refer to the material in the radiating bow shocks; densities (and temperatures)
between the bow shocks are likely to be lower. The temperature where [S~II] radiates
in the HH~270 bow shocks will be $\sim$ 7000~K. Material in the jet
is mostly H, so using the ideal gas law we obtain 1$\times 10^{-8}$ dyne$\,$cm$^{-2}$ for the pressure.

As shown in Table~1, a $\sim$ $5\times 10^{16}$, b $\sim$ $2\times 10^{-20}$, and
c $\sim$ $3\times 10^{-19}$. With this scaling, 200~ns in the
laboratory experiments corresponds to $\sim$ 80 years for an HH flow.
Bow shocks in HH objects typically move their own diameter in $\sim$ 20 years, 
so the agreement with the observed timescales of HH objects and the experiment
is ideal. However, the jet velocity transforms less well, with 10 km$\,$s$^{-1}$ in the 
experiment scaling to $\sim$ 40 km$\,$s$^{-1}$ in the stellar jet, where the
actual velocity is a factor of four higher. This difference arises in part 
because the stellar jet cools radiatively, lowering the pressure and 
therefore the value of c. Another way to look at the velocity scaling is to
consider the Euler number V/V$_E$, where V$_E$ = (P/$\rho$)$^{0.5}$ is the
sound speed for $\gamma$=1. The Euler number in the HH~270 jet is about
20, and for other stellar jets may range up to 40. In the experiment this
number is only $\sim$ 6.

The main effect of the difference in Euler numbers is that the 
experimental jet has a wider opening angle than the stellar jet does. However,
both numbers are significantly larger than unity, so both jets are
highly supersonic.  As we discuss in section 2.3,
collapse and subsequent rebound of the washer along the axis of the flow
determines to a large extent how the experimental
jet is shaped, and has no obvious astrophysical analog. For this reason we will focus
our analysis primarily on the leading bow shock and on entrainment of material in the obstacle
rather than on the collimation of the jet.

The working surfaces are generally modeled well by the experiment.
In these regions the critical parameters are the timescales, which match very well,
the Mach numbers in the shocks, and the density contrasts between the jet and
the material ahead of the working surface.
In stellar jets, the velocity of the bow shock into the preshock medium (equivalently,
the velocity of the preshock medium into the bow shock) is similar to that
of the jet, $\sim$ 200 km$\,$s$^{-1}$, because the jet is much denser than the ambient medium. 
The sound speed in the ambient medium can range from 1 $-$ 10 km$\,$s$^{-1}$, depending on
how much ambient ultraviolet light heats the preshock gas. So the Mach number of the leading
bow shock can range from about 20 $-$ 200.
In our experiments, the sound speed ahead of the bow shock is low ($\sim$ 0.03 km$\,$s$^{-1}$),
so the Mach number of the leading bow shock in the experiments is also very large, $\sim$ 200.
When a stellar jet encounters a stationary obstacle
like a molecular cloud, the preshock sound speed is that of the molecular cloud,
$\lesssim$ 1 km$\,$s$^{-1}$, so the Mach number of material entering
the shock into the molecular cloud is $\gtrsim$ 200. The Mach number of this shock
in the experiments depends on the impact parameter, but is typically
$\gtrsim$ 100.

The ratio $\eta$ of the density in the jet to that in the ambient medium
to a large extent determines the morphology of the working surface that
accelerates the ambient medium.  Overdense jets with $\eta$ $>$ 1 act like
bullets, and have strong bow shocks and weak Mach disks, while the opposite case
of $\eta$ $<$ 1 produces jets that `splatter' from strong Mach disks (sometimes called
hot spots) and create large backflowing cocoons \citep{krause}. Stellar jets are overdense, with
$\eta$ $\sim$ 10 while extragalactic jets are underdense with $\eta$ $\lesssim$
$10^{-3}$ \citep{krause}.  In our experiments,
$\eta$ ranges from 8 $-$ 1 between 50~ns and 200~ns, respectively, in good
agreement with the overdense stellar jet case.
For the obstacle, the molecular cloud density increases from at least an order of magnitude
less than that of the jet in the periphery of the cloud (essentially the ambient medium
density), to $\gtrsim$ 2 orders of magnitude
larger than the jet at the center of the core. So the equal densities of the jet
and obstacle in the experiment cover a relevant astrophysical
regime.  The main difference between the experimental and astrophysical cases
is the uniform density of the experimental obstacle
as compared with the $\sim$ r$^{-2}$ density falloff in an isothermal molecular cloud core.

Internal shock waves are by far the most common kind of shock in
a stellar jet. Here, velocity perturbations of
30 $-$ 60 km$\,$s$^{-1}$ produce forward and reverse shock waves within a jet
as it flows outward from the source at 150 $-$ 300 km$\,$s$^{-1}$ \citep[e.g.][]{hartigan01}.
Gas behind these shock waves cools rapidly to $\sim$ 4000~K by emitting permitted and forbidden
line radiation \citep[e.g.][]{hrh87}, and then more slowly thereafter.
Typically the temperature falls to $\sim$ 2000~K before the gas encounters
another shock front, so the sound speed of the preshock gas is
$\sim$ 5 km$\,$s$^{-1}$.  Hence, the internal Mach number is $\sim$ 10 for these shock waves.  

Unlike the astrophysical case, our experiments do not produce multiple velocity pulses,
and so are less relevant to studying the internal shocks within the beams of stellar jets.
However, when jets are deflected obliquely from an obstacle the velocities in the deflected
flow can remain supersonic, and may form weaker shocks in the complex working surface area that
lies between the leading bow shock and the deflection shock (Mach disk analog) within 
the jet.  Our experiments are very useful for studying the dynamics of this region, and for
investigating how a jet penetrates and accelerates an obstacle along its path. 

For scaling to hold, several additional conditions must apply. 
First, dissipation mechanisms such as viscosity and heat conduction must
be negligible. Hence, the Reynolds number
$Re \sim VL/(\nu_{mat}+\nu_{rad})$ and Peclet number $Pe \sim VL/\chi$, 
need to be much larger than unity, where $\nu_{mat}$ and $\nu_{rad}$ are kinematic
viscosities for matter and radiation, respectively, and $\chi$ is the
thermal diffusivity.  Using expressions for $\nu_{mat}, \nu_{rad}$, and $\chi$
from \citet{ryutov99} and \citet{drake06}, we find 
thermal conductivity and viscosity are unimportant for both the experimental
and astrophysical cases (Table 1).

The Euler equations above implicitly assume that the gas 
behaves as a polytrope. In reality, stellar jets cool by 
radiating emission lines from shock-heated regions, and so are
neither adiabatic ($\gamma$ = 5/3) nor isothermal ($\gamma$ = 1). 
Existing numerical simulations provide some insight into how jet
morphologies change between the limiting cases of adiabatic and isothermal.
In the adiabatic case, material heated by strong shocks cools by
expanding, so working surfaces of jets tend to have rounder and
more extended bow shocks, while working surfaces
collapse into a dense plug for strongly-cooling jets \citep[e.g.][]{blondin90}.
The effect is less pronounced for weaker oblique shocks like those we are studying
here in the wakes of the deflected flows. However, cooling should
affect the morphologies of the stronger shocks as described above.

In order to use the Euler equations, the laboratory jet must act as a fluid.
Hence, the material mean free path ($\lambda_{mat} \sim
v_{th}/\nu_c$ where $v_{th} \sim \sqrt(k_BT/m)$ is the particle thermal
velocity and $\nu_c$ is the sum of ion and electron collisional frequencies) of
the electrons and ions must be short compared with the size of the system
($\tau_{mat} \sim L/\lambda_{mat} \gg 1$).
Table~1 shows that this condition is easily satisfied
in both stellar jets and the laboratory experiments.

The Euler equations do not include the radiative energy flux, and 
so we must verify that this number is small compared with the hydrodynamical
energy flux.  When, as in our experiments, the minimum radiation mean free path
$\lambda_{rad}$ from Thomson scattering or thermal
bremsstrahlung is short compared to $L$ (i.e., $\tau_{rad}
\sim L/\lambda_{rad} \gg 1$), the
appropriate dimensionless parameter for determining the importance of
radiation is the Boltzmann number Bo\#\ $\sim \rho U^3 / f_{rad}$, where
$U$ is a velocity scale and $f_{rad}$ is the radiative flux.
In the Omega experiments, Bo\#\ $>>$ 1, implying that
the radiative energy flux is unimportant in the flow dynamics.
Radiative fluxes are also negligible in optically thin astrophysical 
shocks like those in stellar jets. Even in the brightest jets the
observed radiative luminosity is $\lesssim$ $10^{-3}$ of the energy of
the bulk flow \cite{bbm81}, so the observations and experiments
are consistent with regard to radiative energy flux.

Finally, stellar jets have magnetic fields while our experiments 
do not. Field strengths within stellar jets are difficult to measure, but
could dominate the flow dynamics close to the acceleration region 
\citep[i.e., V$_{Alfven}$ $<$ V$_{flow}$][]{hartigan07}.
At the larger distances of interest to these experiments,
fields are weaker and act mainly to reduce compression in the postshock
regions of radiatively-cooling flows \citep{morse92}.  

In summary, the experiments are good analogs of jets from young stars.
The Mach numbers of the jet
relative to the ambient medium and to the obstacle are high
in both cases, the density contrasts between the jet and the obstacle are
similar, and the scaled times match almost exactly.
Both systems behave like fluids, and viscosity, thermal conduction, and radiative
fluxes are unimportant to the dynamics in both cases. The main differences are that stellar jets
cool radiatively and the experimental jets do not, stellar
jets may have magnetic fields that are not present in the experimental jets, and 
the density profile within the experimental jet is unlikely to have an astrophysical
analog.

\subsection{Results}%2.3

The complete suite of our experimental radiographs are
available on the web (http:$//$ sparky.rice.edu$/\sim$hartigan$/$LLE\_shots.html).
Because our experiment generates only a single image for each shot, if we wish
to investigate how the flow changes with time or compare experiments that
have different offsets of the ball from the axis of the jet (we refer to this distance
as the `impact parameter' in what follows), we must first quantify 
the degree to which target fabrication affects the radiographs of the flow.
To this end, we obtained several groups of laser shots that had identical
backlighters and delay times. One such set appears in Fig.~2, where 
the overall morphology of the flow and position of the
bow shock is reproduced well between the shots, but 
irregularities appear in the bow shock shape that are specific to each target.
In the case of the Ti backlighter image the bow shock resolves into a collection of
nearly cospatial shells.
The level of difference between the two images at the left of Fig.~2 indicates
a typical variation caused by target fabrication and alignment. 

We varied the backlighter type, exploring V, Fe, Zn, and Ni. Each backlighter
has a different opacity through the material, making it possible to probe the
different depths within the structure of the flow at any given time. For example,
the Ti backlighter image in Fig.~2 has significantly less opacity than that of
the V images.  We also tested four different types of foam, normal RF,
large-pore RF, TPX (poly-4-methyl-1-pentene), and DVB (divinylbenzene),
and found that the neither the foam pore size nor the foam type had
any effect on the results.

Fig.~3 summarizes the results from our experiments. The time sequence at the top
shows how the jet evolves in a uniform medium without an obstacle. As described in
section 2.1, the experiment first accelerates a plug of Ti into the foam, followed
by a secondary jet of Ti that originates primarily from the collapse of the washer.
At 50~ns, the radiograph shows a flat-topped profile that defines the shape of the
plug, and as the plug proceeds into the foam the leading shock becomes bow-shaped.
Irregularities in the shape of the bow shock are similar in size and shape
to those caused by variations in the manufacture of the target (Fig.~2).
By 100~ns the secondary jet has also formed, but it does not become well-collimated
until about 200~ns. At 150~ns the end of the jet has a flute-shape, with a less-dense
interior (see also Fig.~5, below). This shape is not a particularly good analog for a
stellar jet, so we place less emphasis on this area in our analysis.

The second row in Fig.~3 illustrates the principal shock structures that form when
the jet encounters a sperical obstacle (labeled `Ball' in the Figure) along its path. 
In general we expect two shocks to form when a continuous supersonic jet impacts ambient material $-$
a forward bow shock that accelerates the ambient medium, and a reverse shock, sometimes called
a Mach disk, that decelerates the jet. The area between the forward and reverse shocks is
known as the working surface, and within the working surface there is a boundary known as the
contact discontinuity which separates shocked jet material from shocked ambient (or
shocked ball) material.

The two forward bow shocks, one into the ball and the other from the deflected jet into the
foam, are clearly visible in the radiographs. However, the Mach disks for these bow shocks are more
difficult to see in Fig.~3. At 150~ns the Ti plug is the primary driver
of the deflected bow shock, and the RAGE simulations discussed below show that the shocked plug
is located near the head of the bow shock at this time (Fig.~4).
Fig.~5 shows that a disk-shaped area of high temperature material
exists in the shocked plug and at the end of the flute, and it is tempting to associate these
areas of hot gas with the Mach disk. The situation is complicated by the fact that the plug jet is 
impulsive, so one would expect the Mach disk to disappear once material in the plug has passed
through it, though the temperature will remain elevated in this area. 

The last two rows of images in Fig.~3 depict two sequences of radiographs taken at common
times after the deposition of the laser pulse, but with impact parameters that increase from
left to right. As expected, the jet burrows into the ball more when the impact parameter is
smaller, and is deflected more when the impact parameter is larger. In all cases the forward 
bow shock into the ball is quite smooth, with no evidence for any fluid dynamical instabilities.
In contrast, the contact discontinuity between the shocked jet and shocked ball material
is highly structured. As we discuss further in section 5, the irregularity of the contact
discontinuity plays a major role in breaking up the ball.
Images obtained with the Zn backlighter are best for revealing the morphology of the
secondary jet, which at late times appears to fragment into a complex 
filamentary structure in the region of the flute.

The ability to observe a flow from an arbitrary viewing angle is an attractive
feature of the laboratory experiments that is not available to astrophysicists
who study stellar jets. Fig.~6 shows two pairs of identical laser shots, one
with a 150~ns time delay the other with a 200~ns delay, where the viewing angle
changed by 90 degrees within each pair. The outline of the ball is clearly visible through
the deflected jet in the symmetrical view ($\theta$=0) at 150~ns. Discerning the true
nature of the deflected flow is much more difficult in the symmetrical view, where the
radiograph resembles a single, less-collimated bow shock.

\section{Numerical Simulations of the Laboratory Experiments}

The design of these experiments and their post-shot analysis was done
with the RAGE (Radiation Adaptive Grid Eulerian) simulation code. We have also adapted
the astrophysical MHD code AstroBEAR to model laser experiments by 
including a laser drive and real equation of state for different materials, but we
summarize the results of this work elsewhere \citep{carver09}.

\subsection{RAGE Simulations}

RAGE is a multi-dimensional, multi-material Eulerian radiation-hydrodynamics
code developed by Los Alamos National Laboratory and Science Applications
International (SAIC) \citep{gittings08}. RAGE uses a continuous (in space and time)
adaptive-mesh-refinement (CAMR) algorithm to follow interfaces and shocks,
and gradients of physical quantities such as material densities and
temperatures. At each cycle, the code automatically determines whether
to subdivide or recombine Eulerian cells. The user also has the option
to de-zone (that is, reduce the resolution of the mesh) as a function
of time, space, and material. Adjacent square cells may differ by
only one level of resolution, that is, by a factor of 2 in cell size.
The code has several interface-steepening options and easily follows
contact discontinuities with fine zoning at the material interfaces.
This CAMR method speeds calculations by as much as two orders
of magnitude over straight Eulerian methods. RAGE uses a
second-order-accurate Godunov hydrodynamics scheme similar to
the Eulerian scheme of \citet{colella85}. Mixed cells are assumed
to be in pressure and temperature equilibrium, with separate
material and radiation temperatures. The radiation-transfer
equation is solved in the grey, flux-limited-diffusion approximation.

Given the placement of the ball with respect to the symmetry axis
of the undeflected jet, these experiments are inherently three-dimensional.
However, before the jet impacts the ball, the hohlraum and the jet 
are two-dimensional. This allows us to perform highly resolved,
two-dimensional simulations in cylindrically-symmetric geometry to
best capture the ablation of the titanium, the acceleration of
the titanium plug through the vacuum free-run region in the washer,
the collapse of the titanium hole onto the symmetry axis, and
the subsequent jet formation. 

The two-dimensional simulations are initialized by imposing the
measured radiation-drive temperature in a region that is inside
the hohlraum. To save computer time, we have determined that it
is sufficient to eliminate the hohlraum and just use the
measured temperature profile as the source of energy that creates
the ablation pressure to drive the jet.

Neglecting asymmetries that arise from initial perturbations and
unintentional misalignment of the hohlraum with the vacuum
free-run region (the hole in the washer), the early-time
experimental data show that the jet remains cylindrically
symmetric until at least 50 ns. At approximately this time we
link the two-dimensional, cylindrically-symmetric simulation
into three dimensions and add the 1~mm diameter ball at
the different impact parameters. While the two-dimensional
simulations are run with a resolution of 1.5 $\mu$m to 
capture the radiation ablation of the titanium correctly, the
three-dimensional simulations are typically run at lower
resolution, especially during the design of the experiments.

Fig.~4 shows the model density, composition, and expected radiograph for a deflected
jet at 200~ns.  Material labeled as blue (Ti plug) comes from the Ti disk,
while orange (Ti sleeve) originates in the Ti washer and constitutes most of the
secondary jet. The flute-shape of the secondary jet is clear in this figure. 
The areas of greatest interest are the region where the jet is creating
a cavity in the ball, because this shows how jets destroy obstacles and
entrain material, and the region of filaments in the working surface of
the leading bow shock, because these filaments could in principle propagate
downstream as weak shocks like those seen in HH~110.

The processes of jet formation and propagation are illustrated in Fig.~5,
which show `snapshots' of the temperature and density distribution
within the jet taken from a RAGE hydrocode simulation (single choice
of axial position and impact parameter for the polystyrene sphere)
at different times. The primary (plug)
and secondary (`shaped-charge' hole collapse) jets are identified in
Fig.~5, as well as the late-time motion at the titanium-to-foam
interface that results in the formation of an opaque pedestal-like
feature at the base of the jet in the synthetic radiographs. All
these features are clearly discernible in the experimental data.
Fig.~7 shows post-processed, zinc-backlit (9 keV x-ray backlighter energy)
radiographic images from two perpendicular views that represent the
time evolution of the jet, and how it subsequently deflects from the ball,
in the RAGE simulation. 

\subsection{Data Analysis and Comparisons with RAGE Simulations}

The experimental data recording the jet hydrodynamics are in the form of
radiographic images recorded on x-ray-sensitive film. Fig.~3 shows a
composite of experimental radiographs from several laser shots, to
illustrate the formation and deflection of the jets, in a series of
shots where the position and impact parameter of the polystyrene
sphere inevitably vary somewhat from shot to shot. We digitized the
film data with a Perkin-Elmer PDS scanning microdensitometer and
converted the film density to exposure using calibration data
for the Kodak DEF film that recorded the images \citep{henke86}. Ideally,
the pinhole-apertured x-ray backlighting source would provide
spatially uniform illumination of the experimental assembly, although
in practice this is not the case because of laser intensity variations
at the backlighting target and vignetting resulting from the specific
size and shape of the pinhole aperture. Regions of the image resulting
from x-ray transmission through the undisturbed foam (that is,
outside the jet-driven bow shock) provide a means to determine
the uniformity of backlighting intensity, after we allow for
the known x-ray attenuation resulting from the
undisturbed foam. Starting from this measured intensity
distribution we use a polynomial fitting procedure to
infer the unattenuated backlighter intensity that underlies the image
of the jet and bow shock. We divide the backlit image data by
the inferred (unattenuated) backlighter intensity to obtain a
map of x-ray transmission through the experiment. We compare
this x-ray transmission data with post-processed hydrocode
calculations that simulate the experimental radiographs.
Because the absolute variation of backlighter intensity
across the image is small (typically, 10 $-$ 20\% across
the entire image), the polynomial fitting procedure enables the
absolute x-ray transmission to be inferred with second-order accuracy.

We use three principal metrics for comparing the experimental
radiographs with synthetic x-ray images obtained by post-processing of
our hydrocode simulations. These are: (1) the large-scale hydrodynamic
motion (determined by comparison of the positions of the bow shock
in experiment and simulation); (2) the spatial distribution of mass of
the titanium jet material (obtained from the spatial integral of optical
depth throughout the image, or from sub-regions of the image); and (3)
the small-scale structure in the deflected jet (quantified by the
two-dimensional discrete Fourier transform of sub-regions of the image,
and its corresponding power spectral density function of spatial frequency).

In making comparison of the experimental with hydrocode simulation,
two specific points of detail require attention: (i)~each laser-target
assembly has its own specific location of the polystyrene sphere
within the polymer foam cylinder (because of target-to-target variations
arising in fabrication), and (ii)~the angular orientation of the
polymer-foam cylinder attached to the laser-heated hohlraum target
determines the x-ray backlighting line of sight, relative to the
plane in which the polystyrene sphere is displaced from the axis
of the jet. Ideally all laser targets would be identical, and
all backlit images would be recorded orthogonal to the axis of
the target, and either in the plane of radial displacement of
the polystyrene sphere or orthogonal to that plane.

The geometry of the target chamber of the Omega laser facility determines the
possibilities for backlighting orthogonal to the axis of the
experimental assembly, and is well-characterized. However,
to model a specific experiment fully, we must run
three-dimensional hydrocode simulations specific to that experiment
which include target-to-target differences of the fabrication 
assembly, followed by post-processing specific to the backlighter line of
sight used in each experiment (which may differ by up to 10 degrees
from the two preferred directions dictated by the symmetry of the experiment).
The three-dimensional hydrocode simulations are expensive in their
use of computing time and resources, and we therefore make
detailed comparison of the experimental data with simulation for
only a small number of representative cases.

\subsubsection{Large-Scale Hydrodynamics and Bow Shock Position}

Figs.~8 and 9 compare images of the deflected jet at 200 ns
after the onset of radiation drive to the experimental assembly, and
for backlighting lines of sight orthogonal to the plane of jet deflection
(Fig.~8) and within the plane of jet deflection (Fig.~9).
In each case, the data are compared with post-processed simulations
from the RAGE hydrocode. The x-ray backlighting source was the 9.0 keV
resonance line of He-like zinc (Fig.~8) or the 6.7~keV line radiation
of He-like iron (Fig.~9). In each case, we corrected 
for spatial variations of incident backlighter intensity (as described above)
and the images are therefore maps of absolute backlighter transmission through
the experiment. The spatial resolution of both experiments was 15 $\mu$m.
The bow shock in the hydrocarbon foam ahead of the jet is clearly visible,
as is the late-time hydrodynamic behavior of the primary (outflow) jet
(the mushroom-like feature lying off axis, resulting from deflection of
the jet) and the secondary (hole-collapse) jet (the dense, near-axis
stem apparently penetrating the initial position of the polystyrene sphere).
Also evident in the radiographs is the mound-shaped pedestal that arises
from motion of the titanium-to-foam interface following shock transit
across this interface at late time.

In the case of each experiment (the two images were obtained from different
experimental shots) the radial offset of the center of the sphere from the axis
(impact parameter) was close to 350 $\mu$m , and the axial position of the
center of the sphere was close to 920 $\mu$m . A single RAGE simulation
is shown for purposes of comparison, in which the impact parameter was
350 $\mu$m, and the axial position for the sphere 915 $\mu$m. The
spatial resolution of this simulation was 3.1 $\mu$m.

We make quantitative comparisons of experiment and simulation by
comparison of lineouts of x-ray transmission in Fig.~10 and Fig.~11.
The simulation reproduces the 
large-scale hydrodynamics of the experiment (bow-shock position,
formation of the deflected jet, motion of the pedestal and the
creation of a Mach-stem-like feature where it meets the bow shock).
However, the simulation does less well with other features of
the hydrodynamics, including the
``clumpiness'' of the deflected jet material and its proximity
to the bow shock running ahead of the deflected jet, and the
small-scale structure at the interface of the titanium washer
and foam (apparently at the surface of the pedestal feature).

The finely-resolved simulations capture more fine-scale structure $-$
the jet does indeed break up into structure similar to that seen in the
experiment with simulations at 1 $\mu$m spatial resolution, although
at a somewhat later time than observed experimentally.
These small-scale structures
tend to have larger local Mach numbers and are therefore closer
to the bow shock. The filigree structure at the pedestal probably
arises from Richtmyer-Meshkov growth of small-scale
machining or polishing marks on the surface of the titanium washer,
following shock transit across this interface. There are also
multiple shock interactions that form Mach stems which are not
captured by these simulations owing to the reduced computational 
resolution in the pedestal area.

\subsubsection{Spatial Distribution of Mass}

To proceed further with quantitative comparison with simulation, we consider the
spatial distribution of mass of the materials present in the experiment. 
At each point in the experimental and synthetic images, the optical
depth at the photon energy of the backlighter radiation is given by

\begin{equation}
\tau = \kappa_1\sigma_1 + \kappa_2\sigma_2 + \kappa_3\sigma_3
\end{equation}

\noindent
where $\tau$ and $\sigma$ are the opacity and areal density (integral of density
along the line of sight) and subscripts distinguish the titanium jet (1),
the hydrocarbon foam (2) and the polystyrene sphere (3) materials, respectively. The
opacity of the titanium jet is significantly greater than that of either
the RF foam or the polystyrene sphere, and 
their temperatures are sufficiently low for the opacity of these
materials to be essentially constant throughout the volume of the experiment.
For example, at 9 keV (the photon energy of the zinc
backlighting source) the opacities of titanium, RF foam and
polystyrene are, respectively, 150, 4.12 and 2.83 cm$^2$g$^{-1}$.

We divide the experimentally measured and simulated images, arbitrarily,
into 500 $\mu$m square regions, and for each region we calculate the
mean optical depth $\bar\tau$ using $\tau$ = $-$ln(I/I$_\circ$), and

\begin{equation}
\bar\tau = {{\int\tau dA}\over{\int dA}}
\end{equation}

\noindent
where I/I$_\circ$ is the measured (or simulated) x-ray transmission,
dA is the pixel area, and the integration extends over the area of each
specific region of the image. A comparison of mean optical depth defined in
this way, for both experiment and simulation, appears in Fig.~12. The
grouping together of adjacent, square regions of the image enables the mean
to be calculated for rather larger areas encompassing all of, or the
majority of, the mass of the deflected jet. In particular, we consider two
larger areas of the images shown in Fig.~12: the rectangular region
composed of six adjacent 500 $\mu$m squares and labeled A, and
the L-shaped region composed of three adjacent 500 $\mu$m squares and
labeled B. These encompass essentially all of the mass of the jet
that has interacted with the polystyrene sphere (region A), and all except the
mass of the primary jet deflected by the sphere (in the case of the
L-shaped region B).

In each case, the simulation reproduces the experimentally
measured optical depth to $\sim$ 10\%: for the rectangular region A,
mean optical depths are 0.94 for the experiment and 0.86 for the simulation;
for the L-shaped region B, mean optical depths are 1.15 for the experiment
and 1.10 for the simulation. The greatest difference between experiment and
the simulation arises in the magnitude and distribution of mass of the primary
(outflow) jet deflected by the obstacle.
Material from this primary jet resides mainly in a further L-shaped
region, labeled C in Fig.~12. Although the mean optical depths
for region C differ by only 15\% (0.72~in the case of experiment, 0.62~in
the case of simulation), the distribution of mass shows significant
variation (mean optical depths of 1.14~and 0.65 in the case of
the 500 $\mu$m square region where the difference is most evident).

Fig.~12 shows that the RF
foam makes only a relatively small contribution to the measured optical
depth, but we may assess the magnitude of this contribution
simply by setting the opacity of the other components of the experiment
(titanium jet and polystyrene sphere) to zero in the post-processing
of the hydrocode simulation. We conclude that for the various
500 $\mu$m square regions of the images shown in Fig.~12, the optical
depth of the foam lies in the range 0.10 $-$ 0.17 (this variation
arises because of chord-length effects, and because density variations
of the foam at the bow shock and within the cocoon). Hence,
RF foam contributes a negligible optical depth
in these experiments.

\subsubsection{Discrete Fourier Transform and Power Spectral Density}

To compare the experimental data with the small-scale
structure of the jet in the simulations, we use the two-dimensional discrete Fourier transform
(DFT) of optical depth. Starting from the experimental (or simulated) images
of the experiment, we obtain maps of optical depth ($\tau$ = -ln(I/I$_\circ$)) and
use the fast Fourier transform algorithm to obtain the DFT of selected regions
of the experimental (or simulated) data and then proceed to calculate the power
spectral density (PSD). Fig.~13 shows an image of the deflected jet within
which we have identified two separate regions (green and dark red boxes), as well as
an area of undisturbed hydrocarbon foam (bright red box), and part of the fiducial grid attached to the surface of
the foam (blue box). For each region, we show the PSD of optical depth. We define the PSD
as the sum of amplitude-squared of all Fourier components whose spatial
frequency lies in the range k $-$ k + dk, where k = (k$_x^2$ + k$_y^2$)$^{0.5}$
and k$_x$ and k$_y$ are orthogonal spatial frequency components of the two-dimensional DFT.

Fig.~13 clearly shows the fundamental spatial frequency of the grid and its
harmonics, the flat spectrum of white noise of the backlighter transmission through the
undisturbed foam, and a spectrum arising from the clumpy, perhaps near
turbulent, structure arising within the deflected titanium jet. Fig.~14
shows an analogous PSD analysis, for three different RAGE simulations of the
experiment. Our approach is the same for the simulation as it is 
for the experimental data: we obtain a map of optical depth from
the simulated radiograph and then the PSD of regions whose size and position
is identical to those chosen in analysis of the experimental data. In the
case of Fig.~14, we show the result of RAGE simulations for three different
levels of AMR calculational resolution: 12.5, 6.25, and 3.125 $\mu$m. Although
the calculational resolution (dimension of the smallest Eulerian cell) differs
in these three cases, we first obtain (by interpolating the
post-processed data) a simulated radiograph with the same spatial resolution
as the experimental data before proceeding to calculate the PSD. This 
procedure enables
us to avoid any potential uncertainty of the scaling of PSD amplitude and
frequency spacing when comparing with the experimental data.

\section{Astronomical Observations and Numerical Models of the
Internal Dynamics within the Deflected Jet HH 110}%4

\subsection{Spectral Maps of HH 110}%4.1

We observed HH 110 on 9 Jan 2008 UT with the echelle
spectrograph on the 4-m Mayall telescope at Kitt-Peak National Observatory
in order to quantify the dynamics present in a shocked, deflected astrophysical jet.
The 79-63 grating and 226-1 cross disperser combined with the T2KB CCD
and 1.5 arcsecond slit gave a spectral resolution of 3.0 pixels, or 11.1 km$\,$s$^{-1}$
as measured from the FWHM of night sky emission lines. The CCD, binned
by two along the spatial direction, produced a plate scale of 0.52 arcseconds
per pixel.  The position angle of the slit was 13 degrees, and the length of the
slit was limited by vignetting to be about 140 arcseconds.
Seeing was 1.3 arcseconds, and skies were stable with light cirrus.
A plot of the slit position superposed on an archival HST image of HH~110 appears
in Fig.~15.

We employed an unusual mode of observation, long slit but with a wide
order separating filter (GG 435). This setup causes orders to overlap
at different spatial positions along the slit, but in the case of HH 110
there is no ambiguity because no continuum sources are present.
The advantage of this setup is that one can obtain position-velocity diagrams
simultaneously for all the bright emission lines, including [S~II] 6716, [S~II] 6731,
[N~II] 6583, [N~II] 6548, and H$\alpha$. The [O~I] 6300 and 6363 lines were also
present, but the signal-to-noise of these faint lines was too low to warrant
any profile analysis.  Distortion in the spectrograph optics causes each emission
line to be imaged in a curved arc whose shape varies between orders. Fortunately,
there is strong background line emission in each of the emission lines from the
HH 110 molecular cloud, and we used this emission to correct for distortion
and to define zero radial velocity for the object. A third degree polynomial matched
the distortion of the night sky emission lines within an rms of about 0.2 pixels
(0.74 km$\,$s$^{-1}$).

It is important to remove night sky emission lines as well as the background emission lines
from the molecular cloud, as these lines contaminate multiple orders in our
instrumental setup. To this end, we imaged blank sky near HH~110 frequently and
subtracted this component from the object spectra after aligning the night sky
emission between the object and sky frames
to account for flexure in the spectrograph. Cosmic rays and hot pixels
were removed using a routine described by \citet{hartigan04}, and flatfielding,
bias correction, and trimming were accomplished using standard IRAF
tasks\footnote{IRAF is distributed by the National Optical Astronomy Observatories,
which are operated by the Association of Universities for Research in
Astronomy Inc., under cooperative agreement with the National Science 
Foundation.}.

Alignment of individual exposures was accomplished in two ways. First, we
compared the position of a star relative to the slit that was captured from
the guide camera, whose plate scale of 0.16 arcseconds per pixel we determined
by measuring the length of the slit image when a decker of known size truncated
the slit. More precise positioning in the direction along the slit is possible by
extracting the spatial H$\alpha$ trace
and performing a spatial cross correlation between frames. The uncertainty in
the positional measurements from the guide camera image inferred from the 
secondary corrections required by the cross-correlations is about 1.0 arcseconds.

To align and compare different emission lines we must also register the
position-velocity diagrams to account for the 
spatial positions of the orders and for the tilt of the spectrum across the
CCD within each order, but this is easy to do with a continuum lamp exposure
through a small decker to define the spectral trace.
The guide camera images indicate, and the spectra confirm, that the last three
object exposures drifted by 1.5 arcseconds relative to the position shown in
Fig.~15, so these were not used in the final analysis. In all, the total 
exposure time for the spectra is 140 minutes. There were no differences between
the [S~II] 6716 and [S~II] 6731 position-velocity diagrams, so we combined these
to produce a single [S~II] spectrum. The [N~II] 6548 line is fainter by a factor
of three than the companion line [N~II] 6583, and the fainter line is also
contaminated by faint residuals from a bright night sky emission line from an
adjacent order, so we simply use the 6583 line for the [N~II] line profiles.

\subsection{Internal Dynamics of HH 110}%4.2

The image in Fig.~15 divides the position-velocity diagram into
ten distinct emitting regions along the slit. Spectra for each of these positions
appear in Fig.~17 for H$\alpha$ and for the sum of [N~II] + [S~II]. All regions
have well-resolved emission line profiles in H$\alpha$, [N~II], and in [S~II].
Only object 9 showed any difference between the [S~II] and [N~II] profiles, with [N~II]
blueshifted by 12 km$\,$s$^{-1}$ relative to [S~II].

In low-excitation shocks like HH~110, the forbidden line emission peaks when the
gas temperature is $\sim$ 8000~K \citep{hartigan95}, which corresponds to
a spread in radial velocity owing to thermal motions of HWHM = ((2ln2)kT/m)$^{0.5}$,
or 2.6 km$\,$s$^{-1}$ for N, and 1.7 km$\,$s$^{-1}$ for S. These thermal line widths
will be unresolved with the Kitt-Peak observations, which have a
spectral resolution of 11 km$\,$s$^{-1}$.  Therefore, the observed widths 
of the forbidden lines measure nonthermal line broadening in the jet. 
This line broadening most likely arises from nonplanar shock geometry or clumpy
morphologies on small scales, as the HST images show structure down
to at least a tenth of an arcsecond.
However, other forms of nonthermal line broadening, such as
magnetic waves or turbulence, may also contribute to the line widths.

Figs.~16 and 17 show that the H$\alpha$ emission line widths are larger than those
of the forbidden lines. This behavior is expected because a component of H$\alpha$
occurs from collisional excitation immediately behind the shock front where the
temperature is highest. The thermal FWHM of H$\alpha$ is given by

\begin{equation}
V_{th} = \left(V_{OBS}^2 - V_{forb}^2\right)^{0.5}
\end{equation}

\noindent
where V$_{forb}$ = (V$_{NT}^2$ + V$_{INST}^2$)$^{0.5}$
is the observed FWHM of the forbidden
lines, V$_{NT}$ the nonthermal FWHM and V$_{INST}$ the instrumental FWHM. 

We can use the thermal line width observation to measure
the shock velocity in the gas. The temperature immediately behind a strong
shock is given by

\begin{equation}
T = {3\over 16}{\mu m_H V_S^2\over k}
\end{equation}

\noindent 
where $\mu$ is the mean molecular weight of the gas and V$_S$ is
the shock velocity. The mean molecular weight depends on the preshock
ionization fraction of the gas, and the postshock gas will have
different ion and electron temperatures immediately behind the shock
until the two fluids equilibrate \citep{mckee74}, but the equilibration distance
should be unresolved for HH~110. For simplicity
we take the preshock gas to be mostly neutral,
so $\mu$ $\sim$ 1.  For thermal motion along the line of sight, the
FWHM of a hydrogen emission line profile is then

\begin{equation}
V_{th} =  2.354\left({k T\over {m_H}}\right)^{0.5}
\end{equation}

\noindent
Combining these two equations we obtain

\begin{equation}
V_S = 0.98 V_{th}
\end{equation}

\noindent
so the shock velocity is closely approximated by the observed thermal FWHM.

Fig.~18 summarizes the kinematics and dynamics within HH~110. The radial velocity 
gradually becomes more negative at distances greater than about $2\times 10^{17}$ cm.
The radial velocities in the different emission lines track one another well.
However, the same is not true for the line widths: both Fig.~17 and Fig.~18 show 
clearly that H$\alpha$ is broader than the forbidden line profiles. This 
behavior is expected because when the preshock gas is neutral, much of the H$\alpha$
comes from collisional excitation immediately behind the shock front where the
temperature is high. The low atomic mass of H also increases its line width
relative to those of N and S. The graph shows that the nonthermal component
of the line profiles stays approximately constant at $\sim$ 40 km$\,$s$^{-1}$, and the
thermal component of H$\alpha$ and the shock velocity are $\sim$ 50 km$\,$s$^{-1}$.

There are two epochs of HST images available (Reipurth PI), separated by about 22 months, and
these data show intriguing and complex proper motions. Regions 1 through 4
have multiple shock fronts some of which move in the direction of the jet while
others move along the deflected flow. Fig.~18 shows that this impact zone
has higher nonthermal line widths than present in the flow downstream, consistent
with the HST data. Throughout this rather broad region, denoted
as such in Fig.~15, jet material impacts the molecular cloud. Hints of this
behavior are evident in the ground based proper motion data of
\citet{lopez05}, which show proper motion vectors directed midway between the
direction of the jet and that of the deflected flow. 

Downstream from region 4, the material all moves in the direction
of the deflected flow and gradually expands in size. The electron density of the
shocked gas declines as the flow expands \citep{ro91}. The slow increase in the
blueshifted radial velocity could arise if the observer sees a concave cavity
that gradually redirects the deflected flow towards our line of sight. One gets
the impression from the HST images of a series of weak bubbles that emerges from
the impact zone. 

Published proper motion measurements \citep{lopez05}
suggest tangential velocities of $\sim$ 150 km$\,$s$^{-1}$
along the deflected flow. Hence, the nonthermal broadening is
$\sim$ 25\% of the bulk flow speed in the deflected jet, while the 
internal shock speeds are typically 30\%\ of the flow speed. The 
temperature immediately behind a 50 km$\,$s$^{-1}$ shock is $\sim$ $5.7\times 10^4$ K,
and will drop to $\lesssim$ 5000~K in areas where forbidden lines
have cooled. The corresponding sound speeds are $\sim$ 20 km$\,$s$^{-1}$ and
8 km$\,$s$^{-1}$, respectively, so the bulk flow speed is $\sim$ Mach 10 in
the deflected flow, while the internal shocks there are $\sim$ Mach 3 $-$ 6.
The magnetosonic Mach numbers will be lower, depending on the field
strength.

\subsection{Wide-Field H$_2$ Images of the HH 110 Region}

In order to better define how the HH~110 jet entrains material from the
molecular cloud, and to verify that the deflected jet model is appropriate for
this object, we obtained new wide-field near-infrared images of the region.
In Figs.~19 and 20 we present a portion of an H$_2$ image taken 21 Sept., 2008
with the NEWFIRM infrared camera attached to the 4-m telescope at Kitt Peak National
Observatory. The image was constructed from 20 individual dithers of 2 minutes apiece
for a total exposure time of 40 minutes. The wide-field image in Fig.~19 shows VLA~1,
the driving source of the HH~110 flow \citep{choi06}. The jet from this source, known
as HH~270, does not radiate in H$_2$ until it strikes the molecular cloud, although
the jet is visible in deep [S~II] images \citep{choi06}.

Our H$_2$ image clearly shows two other jets that emanate
from sources embedded within the molecular cloud core. IRS~1 (aka IRAS 05487+0255)
is a bright near-IR source, while IRS~2 appears to be obscured by a flared disk seen nearly edge-on (as in
HH~30, \citet{burrows96}). These sources drive a molecular outflow along the direction of
the jets we see in the H$_2$ image \citep{ro91}.  Archival Spitzer images of the region at 
mid-infrared (3.6 $\mu$m $-$ 8 $\mu$m), and far-infrared (24 $\mu$m, 70 $\mu$m and 160 $\mu$m)
wavelengths reveal three very bright sources that persist in all bands
in the region, VLA~1, IRS~1, and IRS~2. The spectral energy distribution of IRS~1 
is still rising at 100 $\mu$m, indicative of a heavily embedded source.
The morphology of the HH~30 clone IRS~2 splits into two
pieces at shorter wavelengths in the Spitzer images, consistent with an obscuring disk seen edge-on. Epoch 2000
coordinates for the midpoint of the HH~30-like disk in IRS~2 are 5:51:22.70 +2:56:05, and for IRS~1
are 5:51:22.60 +2:55:43.

The ubiquity of jets in this region is a common occurrence, as most star-forming regions
have multiple sources that drive jets. However, it does potentially bring into
question whether or not what we and others \citep[e.g.][]{rrh96} interpret as a deflected jet
for HH~110 may simply be a distinct flow generated by some other embedded source to the northeast
of the emission. The counter to this argument is that Fig.~19 does not show a source near the apex of
HH 110, and the Spitzer images of the region also show nothing there at mid-infrared wavelengths.
If the hot H$_2$ is dragged out from the molecular cloud core
by the impact of the jet, there should be a spatial offset between the H$_2$ from
the core and the H$\alpha$ in the jet, with the H$_2$ located on the side of the
spray closest to the core. The color composite in Fig.~20 demonstrates this effect
very well, as noted previously for a small portion of the jet
\citep{nc96}.

\section{Connecting the Laboratory Experiments With Astrophysical Jets}%5

Our laboratory results highlight two aspects of the fluid dynamics that
are particularly useful for interpreting astronomical images and
spectra $-$ entrainment of ambient material and the dynamics within contact discontinuities.
The experiments also serve as a reminder that viewing angle affects how
a bow shock appears in an image. We discuss each of these ideas below.

\subsection{Interpretation of Astronomical Images}

Keeping in mind that magnetic fields may play some role in the dynamics, we
can look to the experiments as a guide to how material from an obstacle 
like a molecular cloud core becomes entrained by a jet 
in a glancing collision.  As noted above,
in the astronomical images the H$_2$ emission in Fig.~20 must arise from the cloud
core to be consistent with the observed spatial offset of the H$\alpha$ and H$_2$, and the
lack of H$_2$ in the jet before it strikes the cloud. 
In the experiments, the jet entrains material in the ball in part because
the flute penetrates into the ball and `scoops up' whatever material falls
within the flute (Fig.~4). This type of entrainment is probably of little interest 
astrophysically because its origin is unique to the relatively hollow density structure within the experimental jet.
However, material from the ball is also lifted into the flow because the jet penetrates into
the ball along the contact discontinuity. Both the radiographs and the RAGE simulations
show this region to be highly structured (Figs.~3-5). The experiments indicate that
once a part of the jet becomes deflected into the ball along the contact discontinuity it 
creates a small cavity where the jet material lifts small fragments of the ball
into the flow.  We see the same morphology in the astrophysical case. This
type of entrainment is one that occurs as a
natural outgrowth of the complex 3-D structure along a contact discontinuity. 

As Fig.~6 shows, the deflected bow shock appears much wider when the bow shock
has a significant component along the line of sight. While this result is 
rather elementary, Fig.~6 provides a graphic example that is
important to keep in mind when interpreting astronomical images of bow shocks
when the shocked gas has a large redshift or blueshift. In such cases a flow
may appear to be much less collimated than if one were to observe the bow
shock perpendicular to its direction of motion. 
An example of a wide bow shock oriented at a large angle from the plane of
the sky is HH~32A, which has been thoroughly studied at optical wavelengths 
\citep{beck04}. 

\subsection{Filamentary Structures in the Experiment and the Astronomical Observations}

The working surface of the deflected bow shock in the experiment exhibits an
intriguing filamentary structure in the observed radiographs and in the RAGE
simulations (Figs.~3 and 21) that resembles the filamentary shock waves seen in
HH~110 (Fig.~15).  If differential motions between adjacent filaments
in the working surface are supersonic, then weak shocks like those seen in HH~110 could form
as the filaments interact at later times. 

To test this idea, we generated synthetic images of the Mach number and the velocity along a plane
that contains the center of the ball in Fig.~21. Together, these two images indicate whether shock
waves are likely to form within the working surface at later times. Neither image
alone provides this information: a constant velocity flow with two adjecent
regions of different temperature will show markedly different Mach numbers in
close proximity but will not create a shock; similarly, a shock will only form
between adjacent fluid elements with differing velocities and the same temperature if the 
difference between the Mach numbers exceeds unity.

Velocity differences between the filaments in the working surface in Fig.~21 are typically
1 $-$ 2 km/s, or only about 10\%\ of the initial jet velocity. In contrast, the shock waves
in HH~110 have higher velocities, about 30\%\ of the jet speed (section 4.2; Fig.~18).
Moreover, Mach numbers in the filaments range from $\sim$ 1 $-$ 2, so differences
between the filaments are $\lesssim$ 1, indicating
that the relative motions between the filaments are subsonic. We conclude that the velocity
differences between the filaments in the working surface are too low to form shock waves
unless the filaments cool. Even if that were to occur, the shock velocities will be about
a factor of three smaller relative to the jet speed than those seen in HH~110. Hence,
the best explanation for the shocks observed in HH~110 is that the
source is impulsive, where each pulse impacts the molecular cloud in a similar manner
to our experimental setup.

It is instructive to consider what one might observe in a position-velocity diagram
in the experiment, were it possible to align a slit down the axis of the deflected flow
and obtain a velocity-resolved spectrum, as is possible with astronomical observations.
We show this exercise in Fig.~22, where we have simply assumed the emissivity of the
gas to be proportional to its density. For a real emission line the emissivity depends
on the temperature, density, and ionization state of the gas in a complex manner
determined by the atomic physics of the line. However, it is still instructive to see what
sort of morphologies appear in a p-v diagram of this sort.

The synthetic p-v diagram in Fig.~22 contains a series of arcs that resemble those present
in astronomical slit spectra of spatially-resolved bow shocks \citep{raga86,hrm90}.
These arcs occur when the slit crosses the jet or an interface such as
the cavity evacuated by the jet. In the case of spatially resolved bow shocks, arcs in the 
p-v diagram result from the motion of a curved shell of material, where the orientation of
the velocity vector relative to the observer changes along the slit.
The lesson from the experiment seems to be that in highly structured flows like our deflected jet,
curved cavities and filaments naturally produce p-v diagrams that contain
multiple arcuate features. As in the case of resolved bow shocks, the velocity amplitude of
the arcs in the p-v diagram is on the order of the shock velocity responsible for
the arc.

In an optically thin astrophysical nebula one can measure five out of the six phase space 
dimensions: x and y position on the sky from images, proper motion velocities V$_x$ and V$_y$ from 
images taken at two times, and the radial velocity V$_z$ along the line of sight from spectroscopy.
As we have shown above, V$_z$ is an extremely useful diagnostic of the dynamics of a flow, but
one that is currently not possible to measure in laboratory experiments. If such a diagnostic
instrument could be developed it would open up a wide range of possibilities for new
studies of supersonic flows.

\section{Summary and Future Work}%6

The combination of experimental, numerical and astronomical observational data 
from this study demonstrates the potential of the emerging field of laboratory astrophysics.
In this paper we have studied how a supersonic jet behaves with time as 
it deflects from an obstacle situated at various distances from the axis of the jet.
The laboratory analog of this phenomenon scales very well to the astrophysical
case of a stellar jet which deflects from a molecular cloud core. 
An important component to our study was to 
expand the observational database of best astrophysical
example, HH~110, by obtaining new spatially-resolved high-spectral resolution observations
capable of distinguishing thermal motions from turbulent motions, and by acquiring
a new deep infrared H$_2$ image that can be compared with existing optical
emission line images from the Hubble Space Telescope.

The laboratory experiments span a range of
times, spatial offsets between the axis of the jet and the center of the ball (impact
parameters), viewing angles, opacities (backlighters), and materials.
The experiments are reproducible and do not depend on composition or structure
of foam, or on the pinhole diameter of the backlighter (spatial resolution of the
experiment). Synthetic radiographs of the
experiment from RAGE match the experimental data extremely well, both qualitatively
as images and quantiatively with Fourier analysis. In fact, the agreement is so good that we
have used the synthetic velocity maps from RAGE to compare the internal dynamics 
of the experiment with those that we measure from the new spectral maps of HH 110.

A new wide-field H$_2$ image supports a scenario where HH~110 represents the
shocked `spray' that results from a glancing collision of the HH~270 jet
with a molecular cloud core.  The H$_2$ in HH~110 is offset from
the H$\alpha$ toward the side closest to the molecular core, consistent with
the deflected jet model. The H$_2$ images also uncovered two sources within the core
that drive collimated jets, one a bright near-infrared source, and the other
a highly-obscured source that appears to be a dense
protostellar disk observed nearly edge-on, as in HH~30.

The experiments provide several important insights into how deflected supersonic
jets like HH~110 behave. In the experiment, entrainment of material in the obstacle 
occurs in part because the morphology of the contact discontinuity between the
shocked jet and shocked obstacle easily develops a complex 3-D structure of
cavities that enables the jet to isolate clumps of obstacle material and entrain
them into the flow. A similar process likely operates in HH~110.
The experiments also reveal filamentary structure in the working surface area of the deflected
bow shock, but the relative motion between these filaments is subsonic. Hence, while this
dynamical process will generate density fluctuations in the outflowing gas, it cannot
produce the filamentary structure and $\sim$ Mach 5 shocks shown by the new velocity maps
of HH~110. For this reason the best model for HH~110 remains that of a pulsed jet which
interacts with a molecular cloud core. 

Synthetic position-velocity maps along the deflected jet from the RAGE simulations
of the experiments appear as a series of arcs, similar to those observed in 
astronomical observations of resolved bow shocks. A close examination of the
experimental data shows that these arcs correspond to regions where the slit
crosses different regions of the flow, such as cavities evacuated by the jet, the jet itself, or
entrained material from the ball. This correspondance between the appearance of a p-v diagram
and the actual morphology of a complex flow is an intuitive, although perhaps
unexpected result of studying the dynamics within the experimental flow.

Finally, observations of the deflected bow shock from different viewing angles
emphasize that the observed morphology and collimation properties of bow shocks
depend strongly upon the orientation of the flow with respect to the observer.
As one would expect, a bow shock deflected toward the observer appears less collimated
than one that is redirected into the plane of the sky. The impact parameter
of the jet and obstacle determines how much the jet deflects from the obstacle and
how rapidly the obstacle becomes disrupted by the jet.

While experimental analogs of astrophysical jets are highly unlikely to ever reproduce 
accurate emission line maps, laser experiments can provide valuable insights into how
the dynamics of complex flows behave. Our study of a deflected supersonic jet is only 
one example of how the fields of astrophysics, numerical computation, and laboratory laser
experiments can compliment one another. We are currently embarking on a similar program to
study the dynamics within supersonic flows that are highly clumpy, and other
investigations are underway related to the launching and collimation of jets
\citep{bellan09}.

\acknowledgements
We are grateful to Dean Jorgensen, Optimation Inc., Burr-Free Micro Hole Division,
6803 South 400 West, Midvale, Utah 84047, USA, for supplying the precision-machined
titanium-alloy components for these experiments.
We thank K.~Dannenberg for her assistance in manufacturing the initial targets for
these experiments, the staff of General Atomics for their dedication in 
developing new targets and delivering them on time, the staff
at Omega for their efficient operation of the laser facility, and an anonymous referee
for useful comments regarding scaling.
This research was made possible by a DOE grant from NNSA as part of the NLUF programs
DE-PS52-08NA28649 and DE-FG52-07NA28056.

\begin{center}
\begin{deluxetable}{lcc}
\singlespace
\tablenum{1}
\tablewidth{0pt}
\tablecolumns{3}
\tabcolsep = 0.08in
\parindent=0em
\tablecaption{}
\tablehead{
\colhead{Fluid Parameter} & \colhead{HH 110} & \colhead{Experiment} \\
}
\startdata
L (cm)  [jet size]                   & 1e15   &  0.02  \\
V (km$\,$s$^{-1}$) [jet velocity]     & 150    &  10    \\
$\rho$ (g$\,$cm$^{-3}$) [jet density] & 2e-20  &  1  \\
P (dyne cm$^{-2}$) [jet pressure] & 1e-8  & 3e10 \\
t$_{flow}$ (sec) [flow timescale] & 3e9    &  1e-7  \\
Composition                       & H      &  Ti    \\
T (eV) [temperature of wake]      & 0.6      &  1     \\
Euler number in jet               &22        & 6      \\
$\nu_{mat}$  (cm$^2$s$^{-1}$) [viscosity] & 9e13   &  2e-5   \\
$\nu_{rad}$  (cm$^2$s$^{-1}$) [viscosity] & 2e21   &  1e-12  \\
$\nu_c$  (s$^{-1}$) [collision freq] & 0.002   &  2e13 \\
$\chi$ (cm$^2$s$^{-1}$) [diffusivity] & 5e24 &  5e-18  \\
Re$_{mat}$  [Reynolds number]       & 7e8    &  1e9   \\
Pe$_{mat}$  [Peclet number]         & 1e7    &  3e4    \\
$\lambda_{mat}$ (cm) (mean-free path) & 3e8 &  8e-9 \\
$\lambda_{rad}$ (cm) (mean-free path) & $>>$L$^a$ &  3e-5 \\
$\tau_{mat}$ [optical depth] & 9e6       &  2e6    \\
$\tau_{rad}$ [optical depth] & $<<$1$^a$ &  7e2    \\
Bo\#                               & $>$1e3  &  2e4    \\

\enddata
\tablenotetext{a} {Non-resonance emission lines in stellar jets are optically thin, while
resonance lines such as Ly$\alpha$ are optically thick.}
\end{deluxetable}
\end{center}
\null\vfill\eject

\begin{figure} %fig1=nfig1
\def\thefigure{1}
\includegraphics{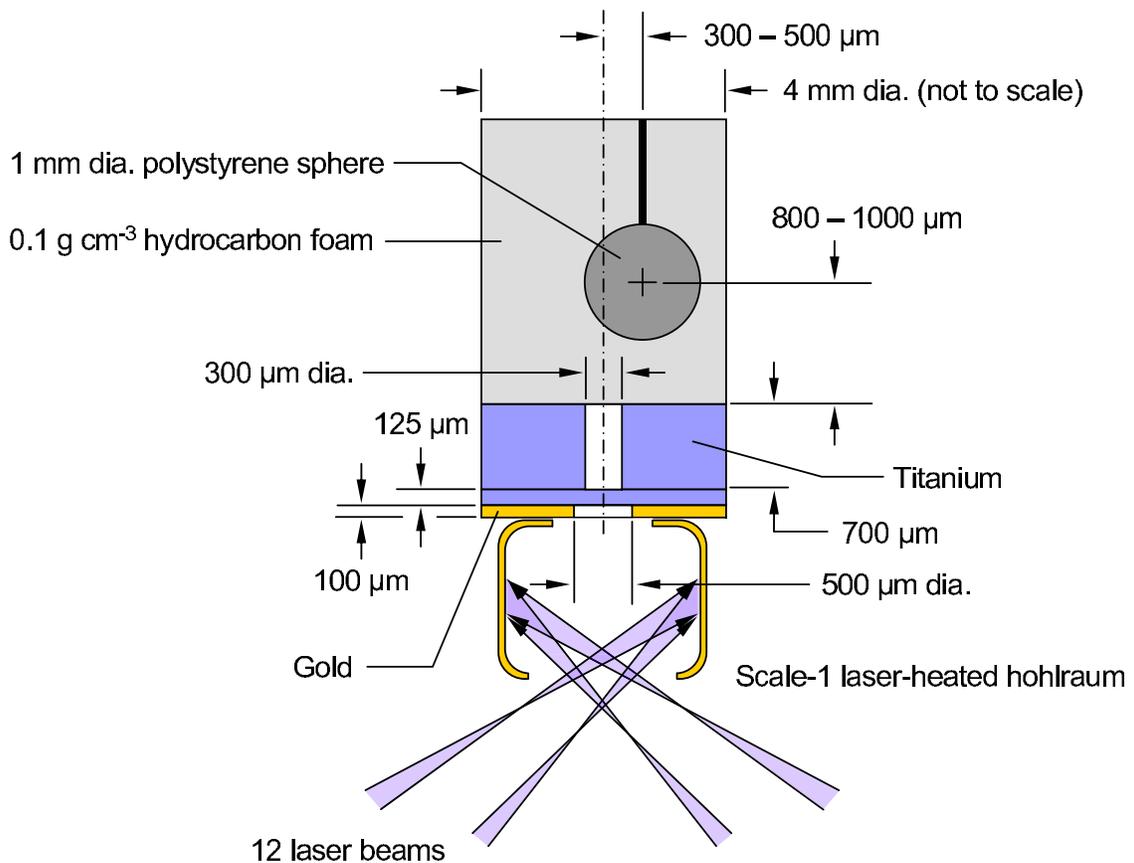}
\caption{Cross section of the target design for the Omega laser experiments.
X-rays produced when the laser impacts the hohlraum accelerate a shock wave
into the Ti disk, and a piece of this disk accelerates up the axis (dashed line)
to produce a supersonic jet. Subsequent collapse of the evacuated cavity generates
a secondary jet along the same direction.  The jets move into a cylinder of hydrocarbon foam, and
deflect from a polystyrene sphere located at some distance (the impact
parameter) from the axis of the cylinder.
}
\end{figure}
\null\vfill\eject

\begin{figure}
\def\thefigure{2}
\includegraphics{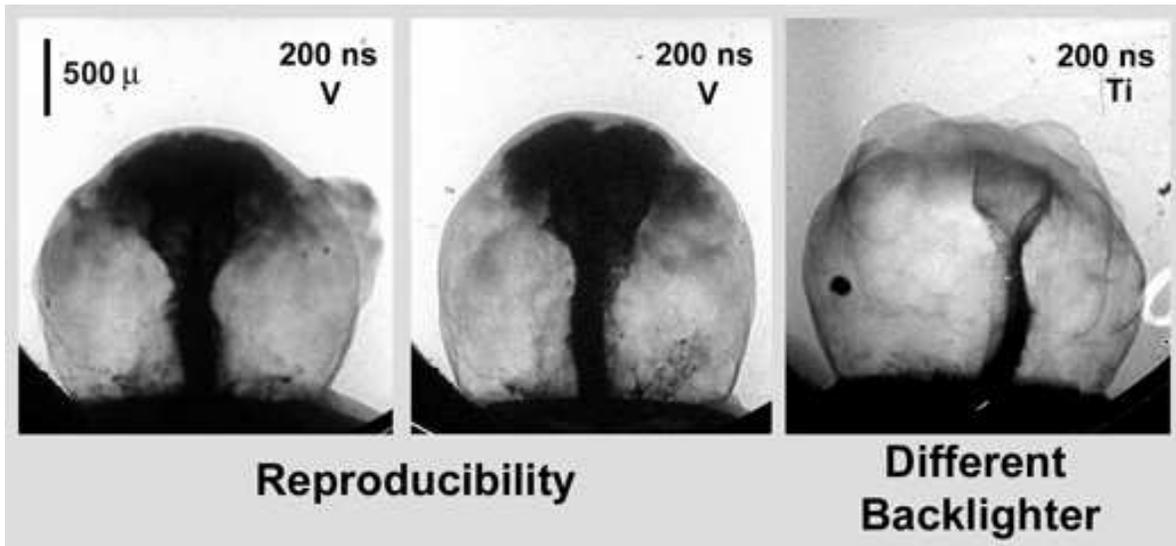}
\caption{Effect of target fabrication, alignment, and backlighter type on
the experiment. Left and center: radiographs of targets without obstacles, taken
at 200~ns with a V backlighter. Small differences in the shape of the
bow shock and jet are caused by target irregularities. Right: Same as the
other images but with a Ti backlighter, which penetrates deeper into the flow.
}
\end{figure}
\null\vfill\eject

\begin{figure}
\def\thefigure{3}
\includegraphics{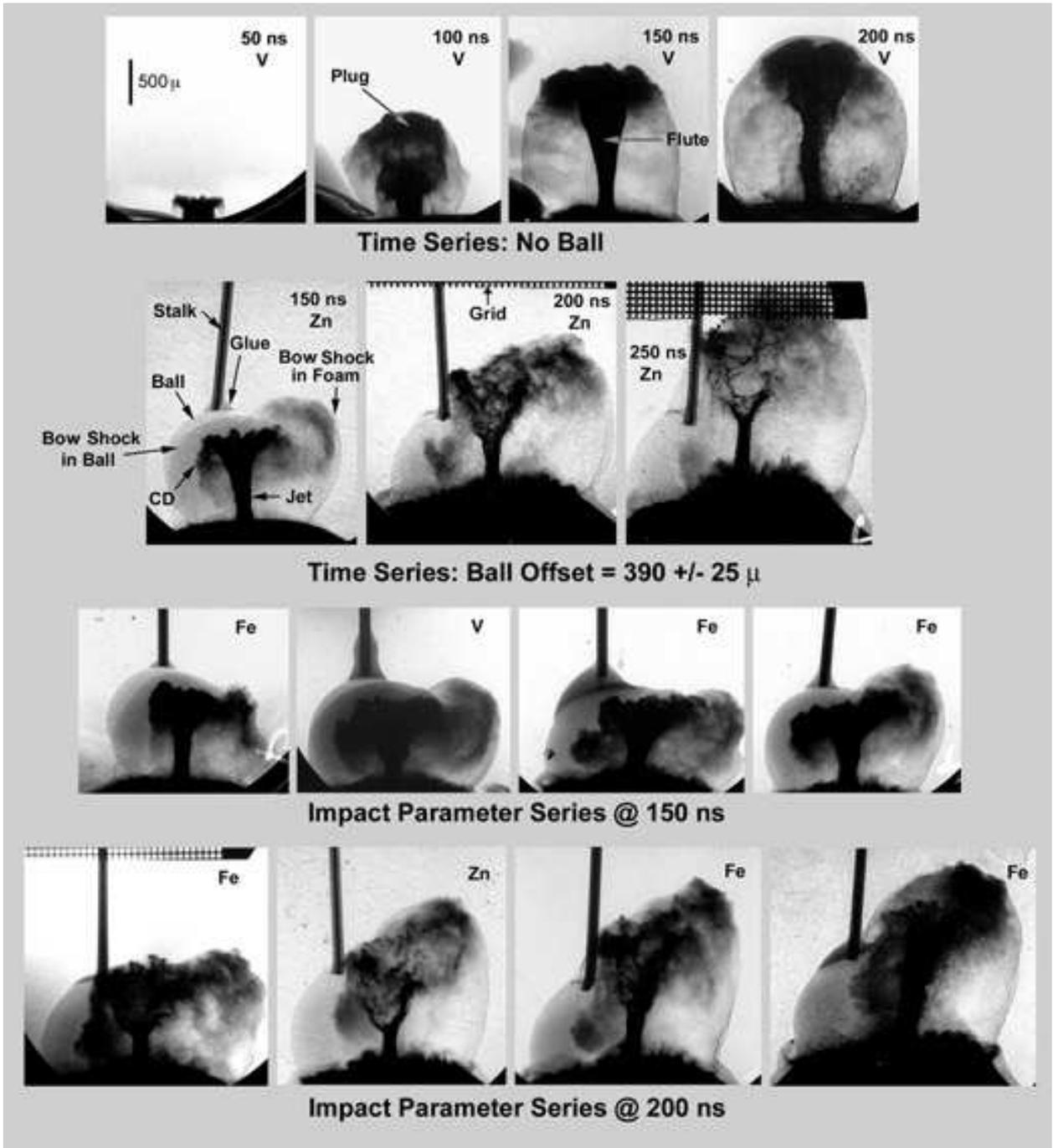}
\caption{Experimental radiographs of deflected supersonic flows. Top: time
series without an obstacle; second row: time series with an obstacle; bottom 
two rows: increasing the impact parameter (left to right) offset of the ball 
from the axis of the jet for delay times of 150~ns and 200~ns. The backlighter
types (V, Zn, or Fe) are shown. CD denotes the contact discontinuity
between the shocked jet and shocked ball material.
}
\end{figure}
\null\vfill\eject

\begin{figure}
\def\thefigure{4}
\includegraphics{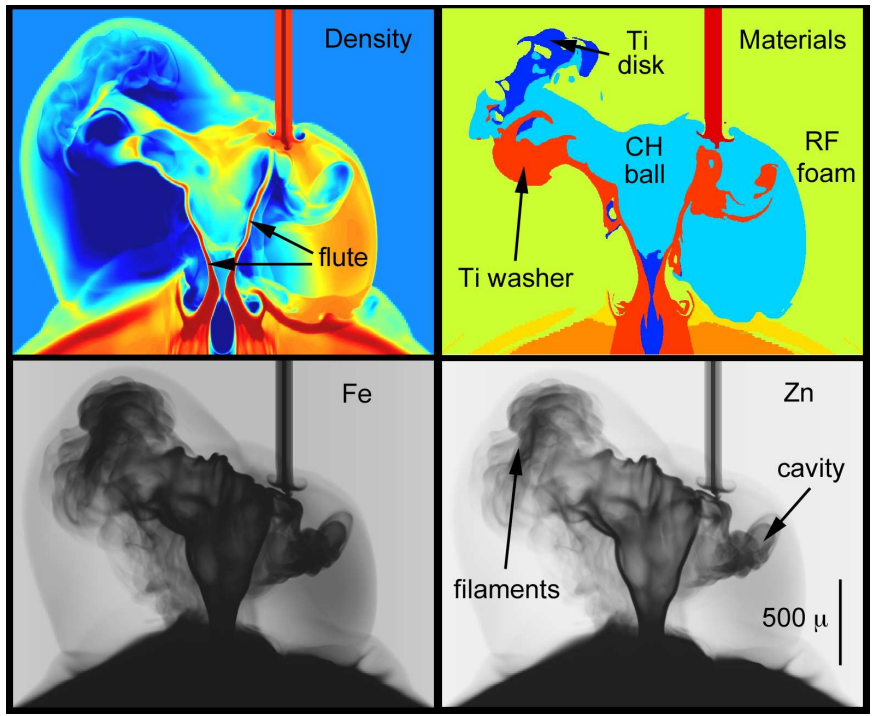}
\caption{RAGE simulations of the deflected flow at 200~ns with an impact
parameter of 350 $\mu$m. Top Left: a slice through the
center of the ball showing density (see also Fig.~5).
Top Right: composition of the flow. Dark blue
denotes material that originated in the Ti disk and was accelerated as 
the plug, orange represents material from the Ti washer that emerged as
the secondary jet. Other materials are
the CH ball (light blue), RF foam (yellow-green), and stalk (red). Bottom: simulated radiographs 
using an Fe and a Zn backlighter.  The filamentary structure of the working surface and the cavity
driven into the ball by the jet are discussed in the text.
}
\end{figure}

\begin{figure} %fig5 = nfig5
\def\thefigure{5}
\includegraphics{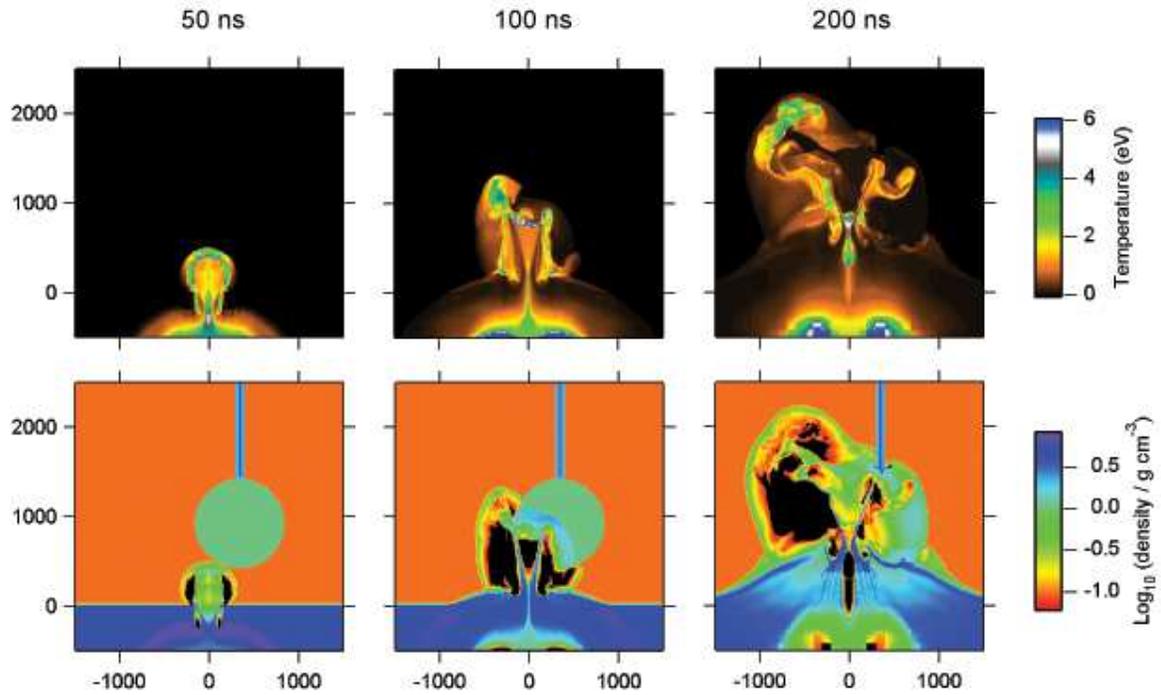}
\caption{RAGE simulations of the laboratory experiments. The top and
bottom rows show the temperature and the density, respectively,
at three different times.  Scales on the x- and y-axes of the images are in
microns.
}
\end{figure}

\begin{figure}
\def\thefigure{6}
\includegraphics{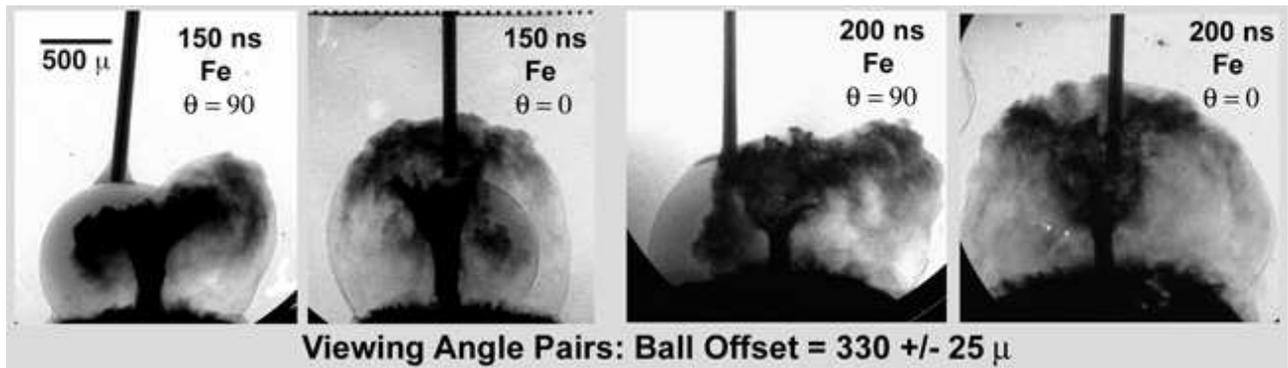}
\caption{Radiographs of a pair of nearly identical targets imaged
with the same backlighter and delay time with orthogonal orientations
at 150~ns (left) and 200~ns (right). The ball is clearly visible 
through the jet in the symmetrical view ($\theta$ = 0) at 150~ns.
}
\end{figure}
\null\vfill\eject

\begin{figure} %fig3=nfig7
\def\thefigure{7}
\includegraphics{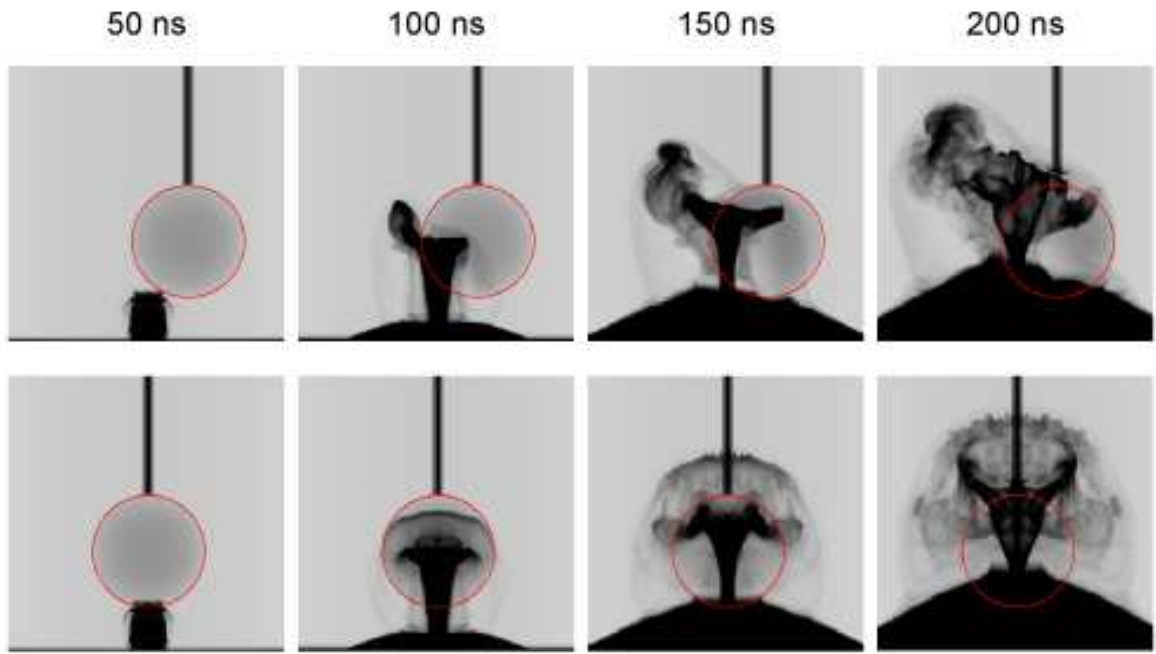}
\caption{Same simulation as in Fig.~5 but showing predicted radiographs.
The location of the ball is marked with a red circle.
Top row: asymmetrical view perpendicular to the plane defined by
the axis of the jet and the center of the ball. Bottom row: symmetrical view,
oriented 90 degrees from the asymmetrical view.
}
\end{figure}
\null\vfill\eject

\begin{figure} %fig4=nfig8
\def\thefigure{8}
\includegraphics{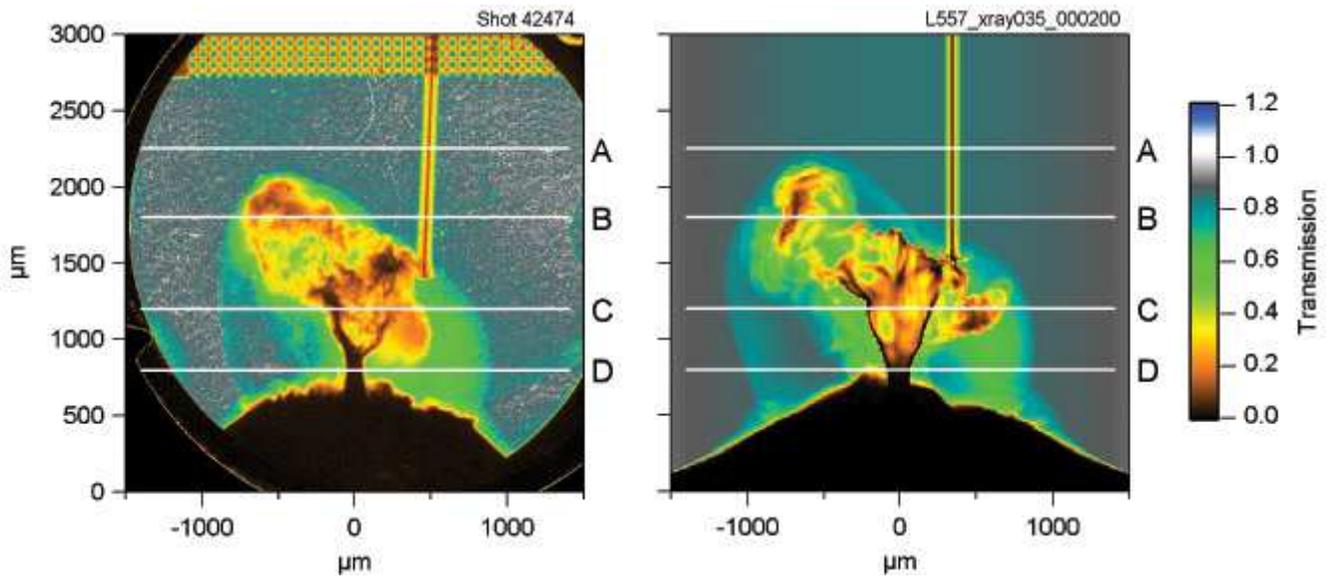}
\caption{Observed (left) and RAGE simulation (right) of a radiograph image
of a deflected supersonic jet. The horizontal lines marked A, B, C, and D mark
locations used for further analysis (cf. Figs.~10 and 11).
}
\end{figure}
\null\vfill\eject

\begin{figure} %fig5=nfig9
\def\thefigure{9}
\includegraphics{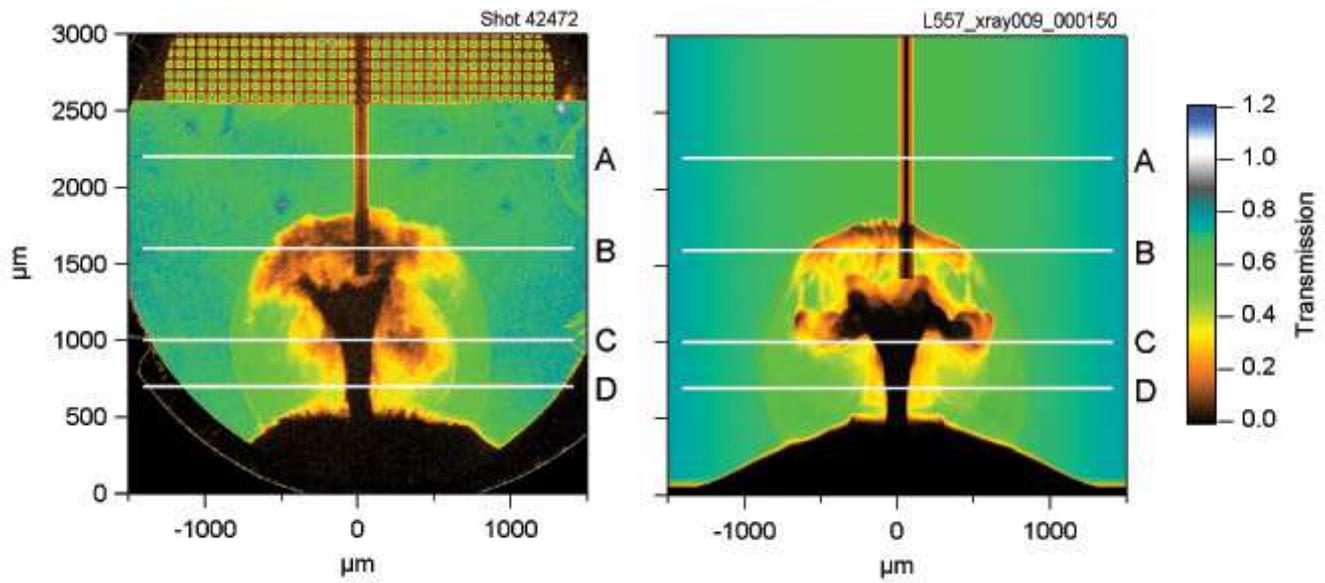}
\caption{Same as in Fig.~8 but for a symmetrical view.
}
\end{figure}
\null\vfill\eject

\begin{figure} %fig6=nfig10
\def\thefigure{10}
\includegraphics{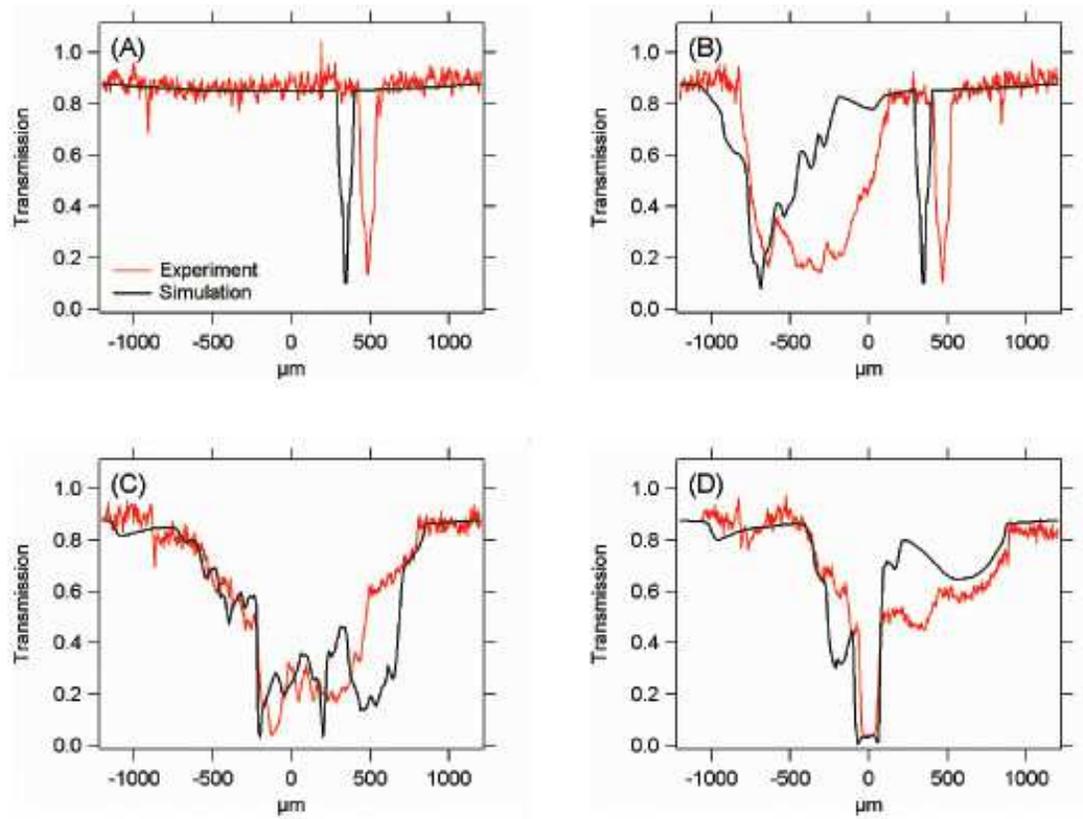}
\caption{Comparisons between the experiment and the simulation along 
the lines marked in Fig.~8.
}
\end{figure}
\null\vfill\eject

\begin{figure} %fig7=nfig11
\def\thefigure{11}
\includegraphics{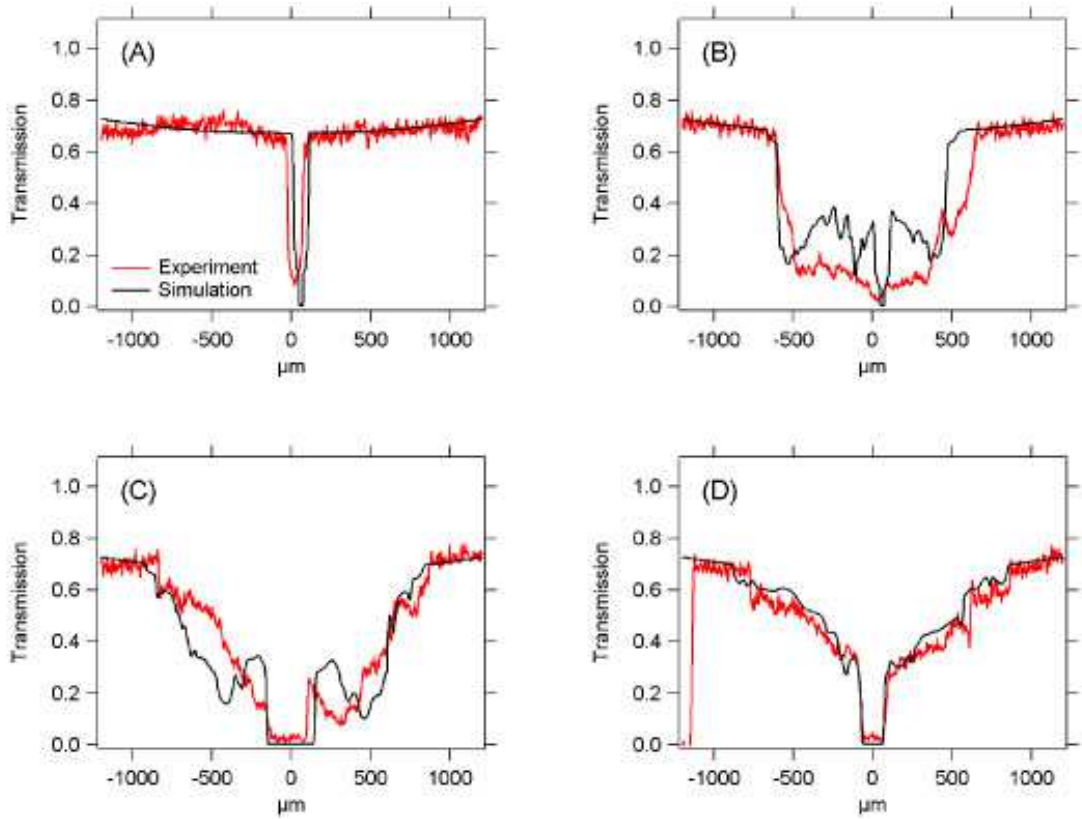}
\caption{Same as Fig.~10 but for the symmetrical view shown in Fig.~9.
}
\end{figure}
\null\vfill\eject

\begin{figure} %fig8=nfig12
\def\thefigure{12}
\includegraphics{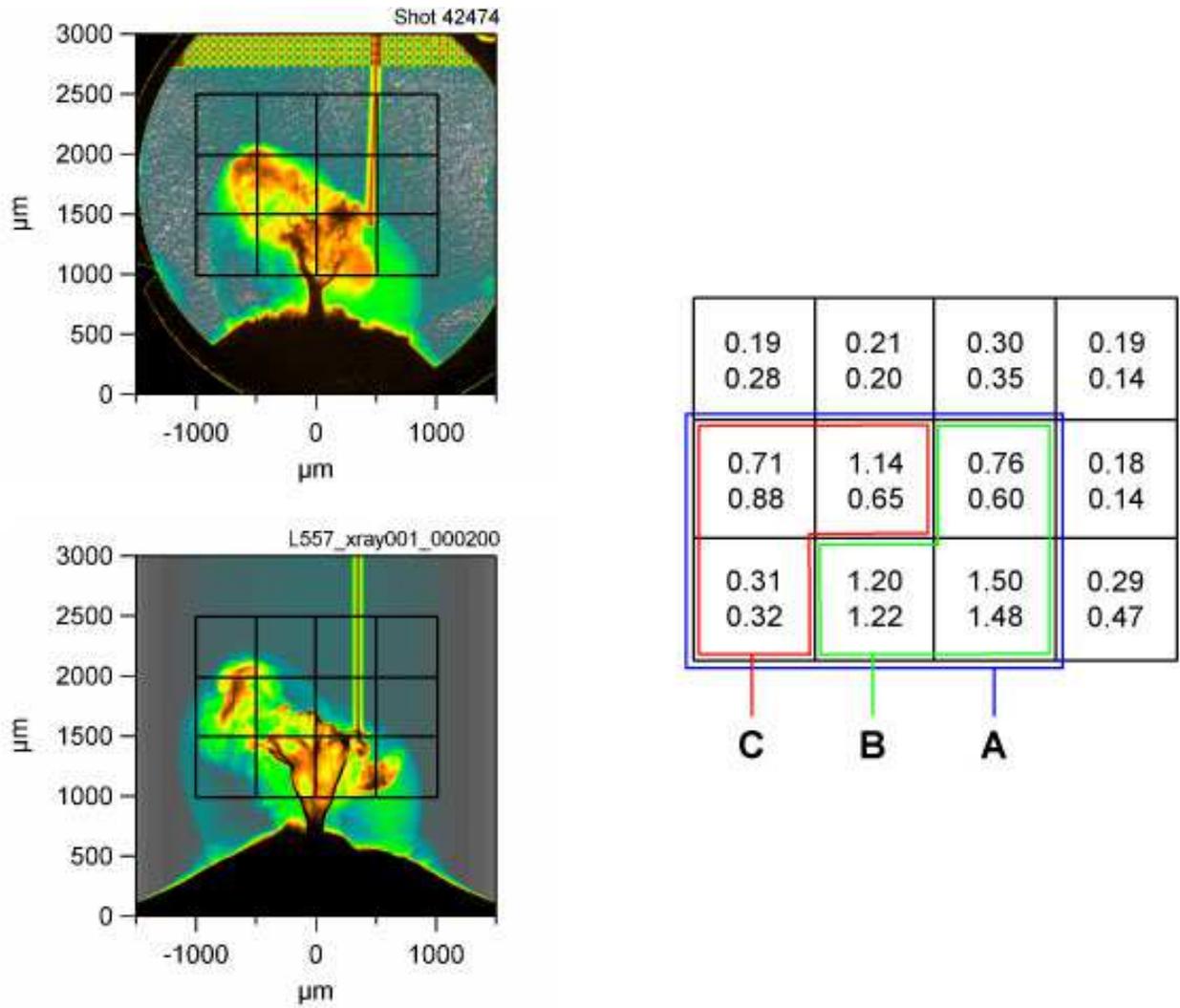}
\caption{Regions defined for more detailed analysis for the observed
radiograph (top) and the RAGE simulation (bottom) as discussed in the
text. The top and bottom numbers in each sqaure are the optical depths
in the experiment, and simulation, respectively.
}
\end{figure}
\null\vfill\eject

\begin{figure} %fig9=nfig13
\def\thefigure{13}
\includegraphics{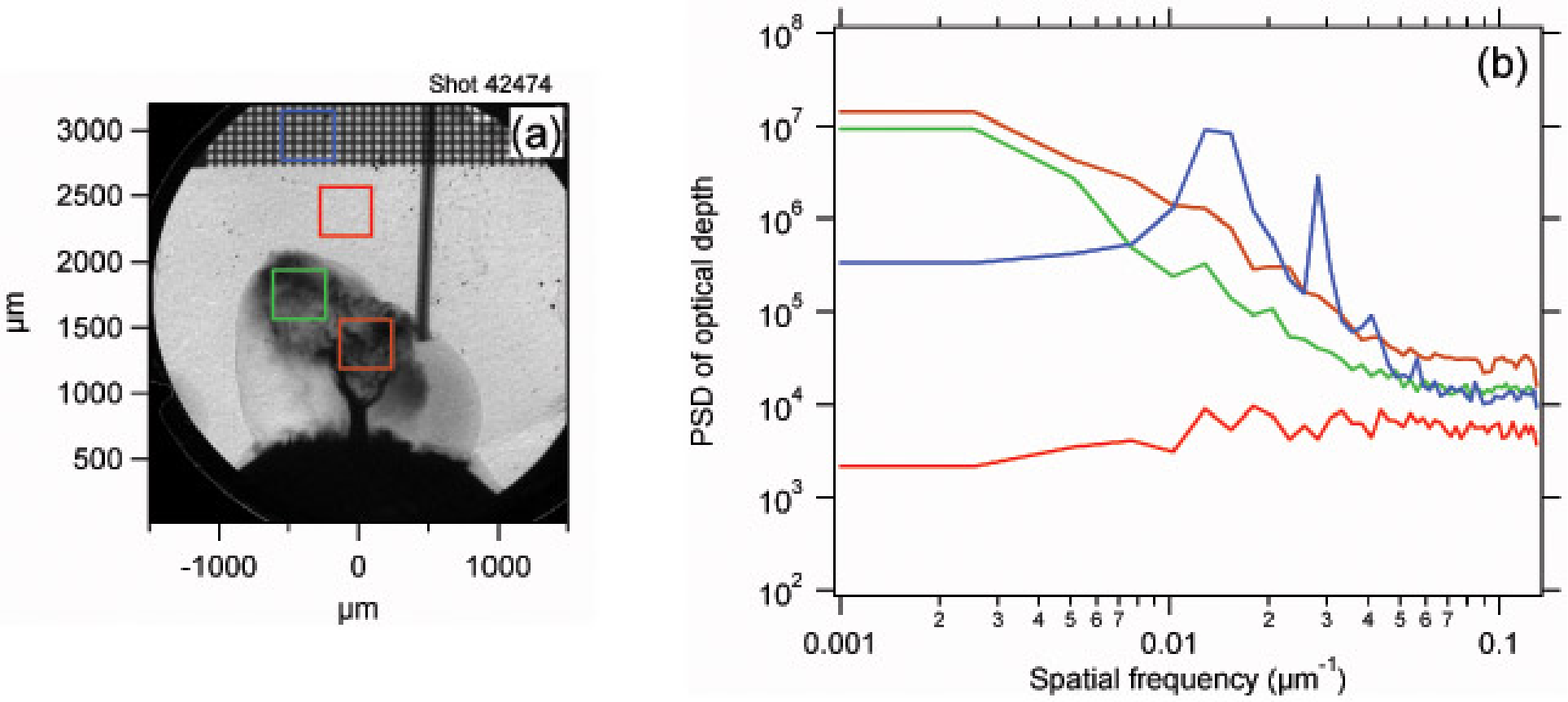}
\caption{Power spectra of discrete Fourier transforms of four regions
in a radiograph image of a deflected jet. The color of the power spectrum
curve corresponds to the color of the box marked in the image.
The periodic structure of the mesh used to define the spatial scale
appears as series of peaks in the blue spectrum. 
}
\end{figure}
\null\vfill\eject

\begin{figure} %fig10=nfig14
\def\thefigure{14}
\includegraphics{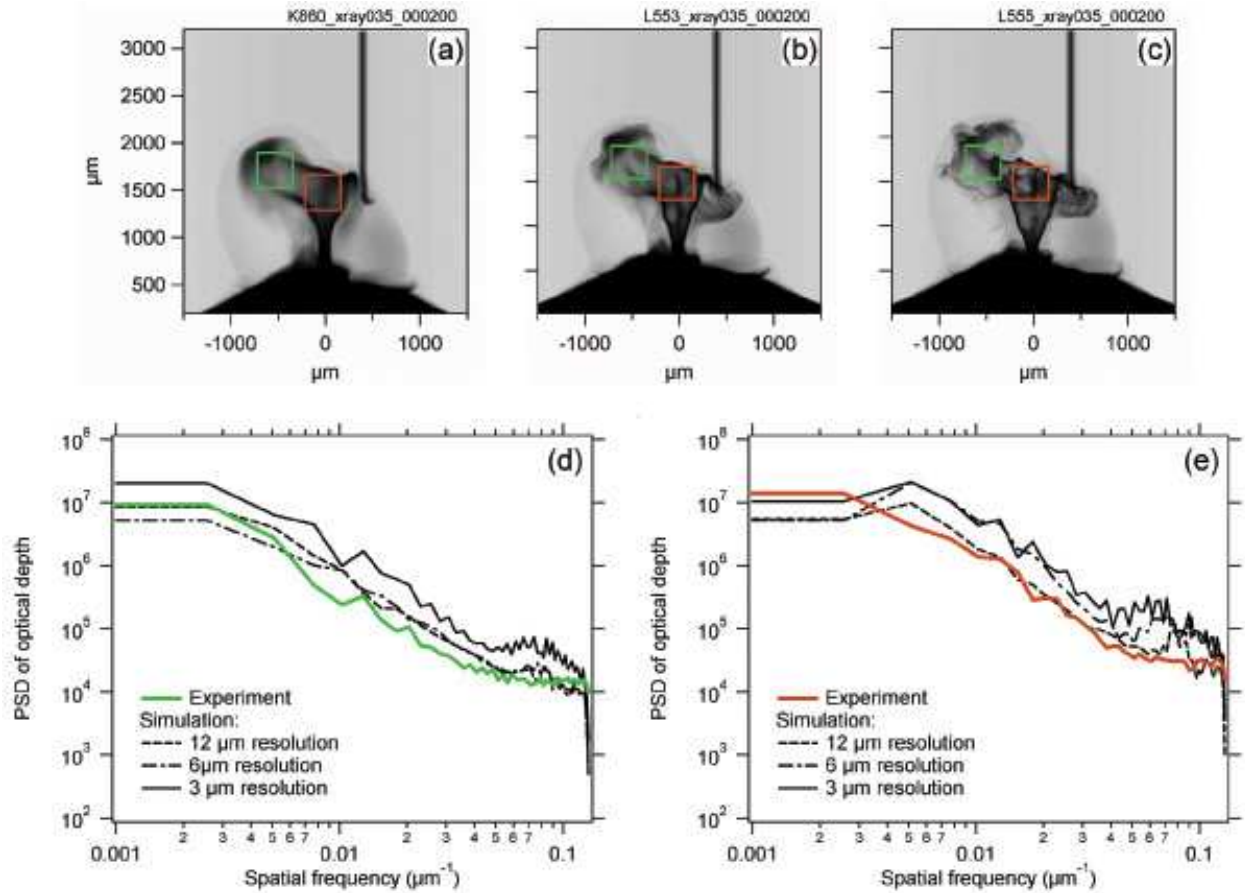}
\caption{Same as Fig.~13 but for RAGE simulations at 12 $\mu$m,
6 $\mu$m, and 3 $\mu$m resolution (panels a, b, and c, respectively).
Panel (d) shows the power spectrum in the deflected wake, while panel (e)
shows a similar spectrum near the axis of the jet (green and red boxes,
respectively). The colored curves in panels (d) and (e) show the corresponding
experimental results.
}
\end{figure}
\null\vfill\eject

\begin{figure} %fig18=nfig15
\def\thefigure{15}
\includegraphics{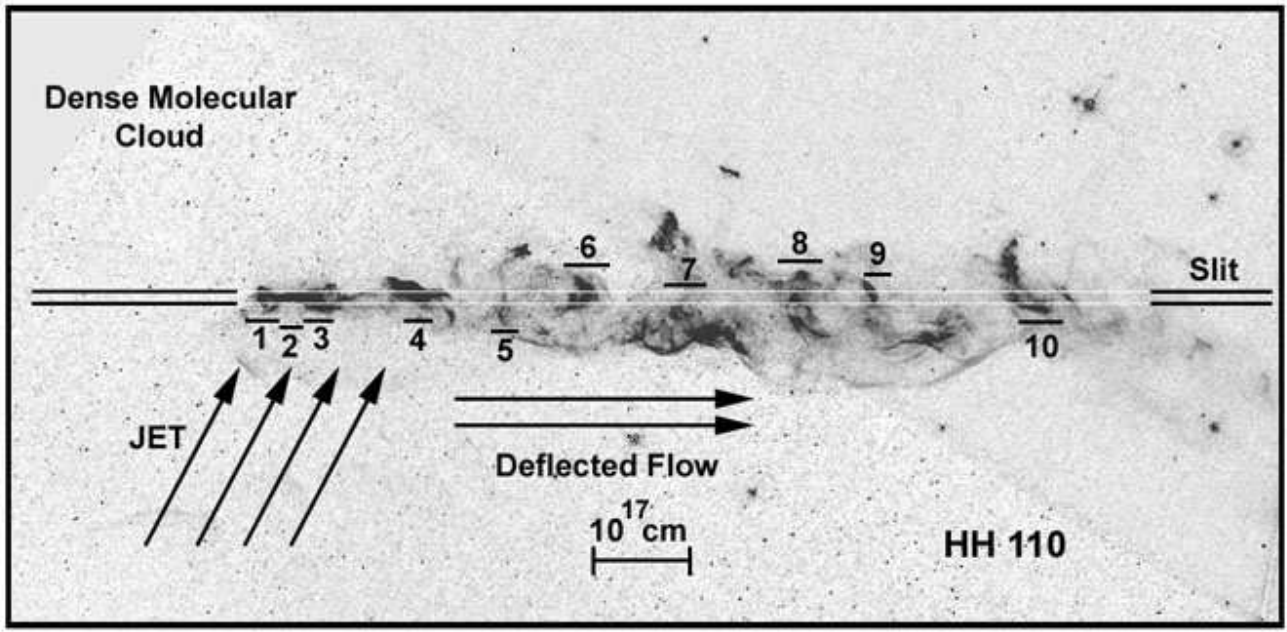}
\caption{
Archival HST emission line image of the HH 110 jet taken through the F658N filter of
ACS (P.I.~Reipurth). The jet deflects from the
molecular cloud as indicated. The slit position used for the spectral
observations is shown, and the numbers mark the ten regions used to extract
line profiles. The spatial scale assumes a distance of 450 pc.
}
\end{figure}
\null\vfill\eject

\begin{figure} %fig19=nfig16
\def\thefigure{16}
\null\includegraphics{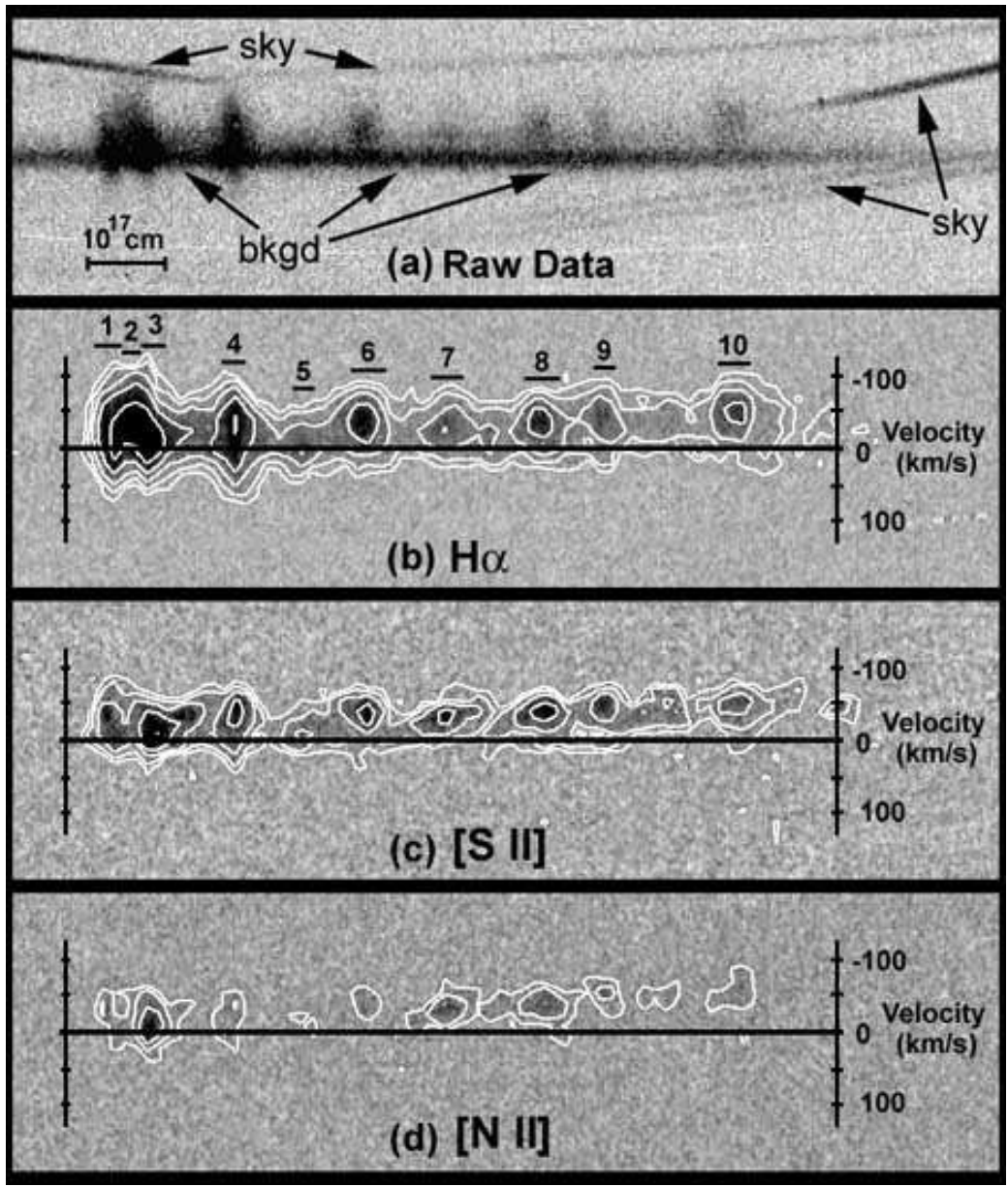}
\caption{Position-velocity diagrams along the HH 110 jet. (a): A typical single
20-minute exposure corrected for distortion but not for background contamination.
Night sky lines from adjacent orders are labeled, and the widths of these lines
shows the instrumental spectral resolution. Emission from the molecular cloud,
denoted as `bkgd', defines zero radial velocity. (b): The H$\alpha$ position-velocity 
diagram, coadded from seven 20-minute exposures and with background subtracted. The
numbered regions correspond to the areas along the slit marked in Fig.~15. The jet
is easily resolved everywhere along the flow. Contours are separated by a factor of
1.41 in flux. (c) and (d): Same as (b) but for [S II] 6716+6731, and [N II] 6583,
respectively.
}
\end{figure}
\null\vfill\eject

\begin{figure} %fig20=nfig17
\def\thefigure{17}
\null\includegraphics{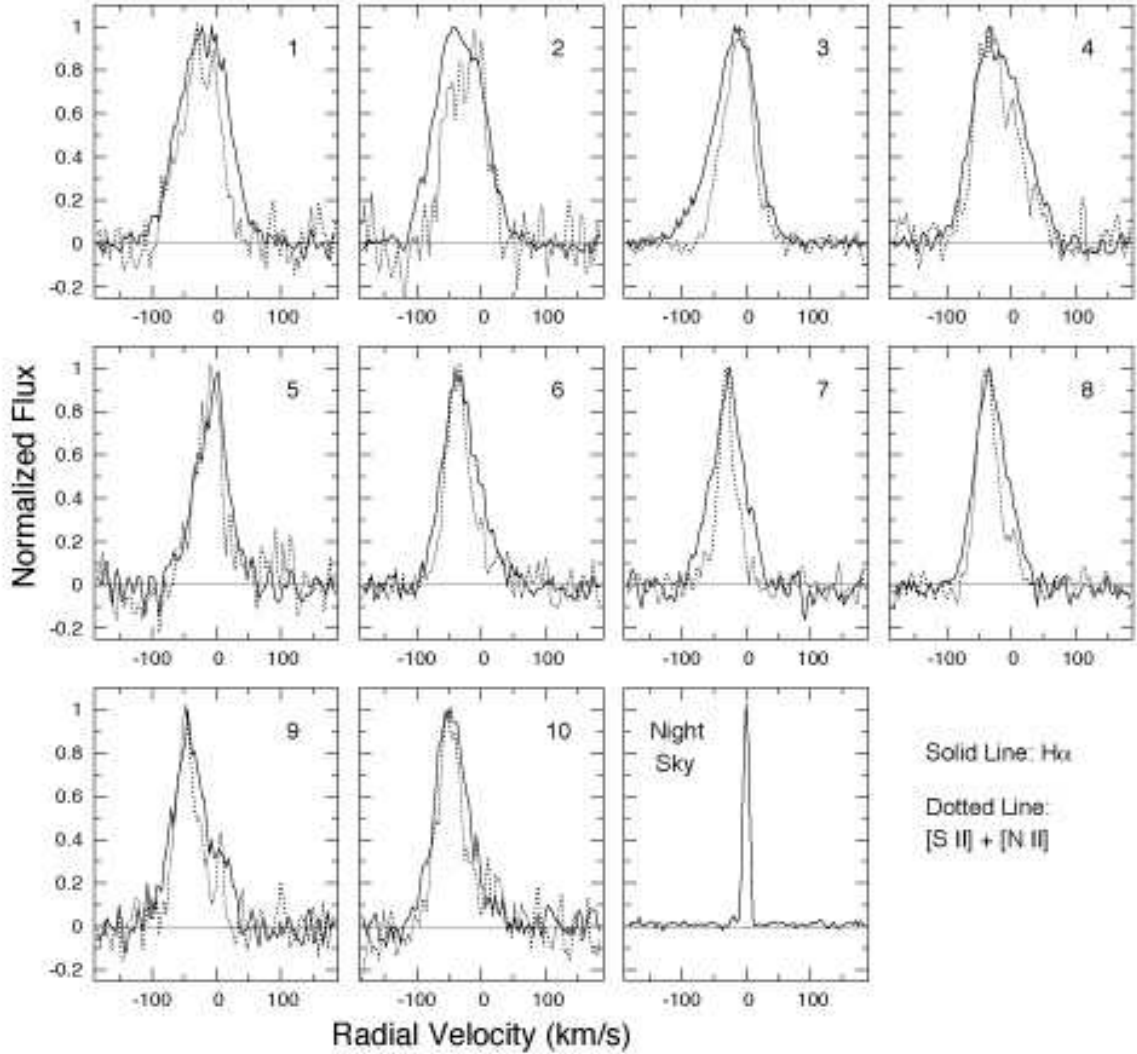}
\caption{Emission line profiles in HH 110. Solid and dotted lines denote the
emission line profiles of H$\alpha$ and [S~II]+[N~II], respectively, for each of
the ten regions along the slit marked in Fig.~15. The night sky profile shows the
instrumental resolution. The jet is resolved everywhere, and is somewhat broader in
H$\alpha$ than it is in the forbidden lines. The radial velocity becomes more 
negative as the jet flows downstream from the impact point.}
\end{figure}
\null\vfill\eject

\begin{figure} %fig21=nfig18
\def\thefigure{18}
\null\includegraphics{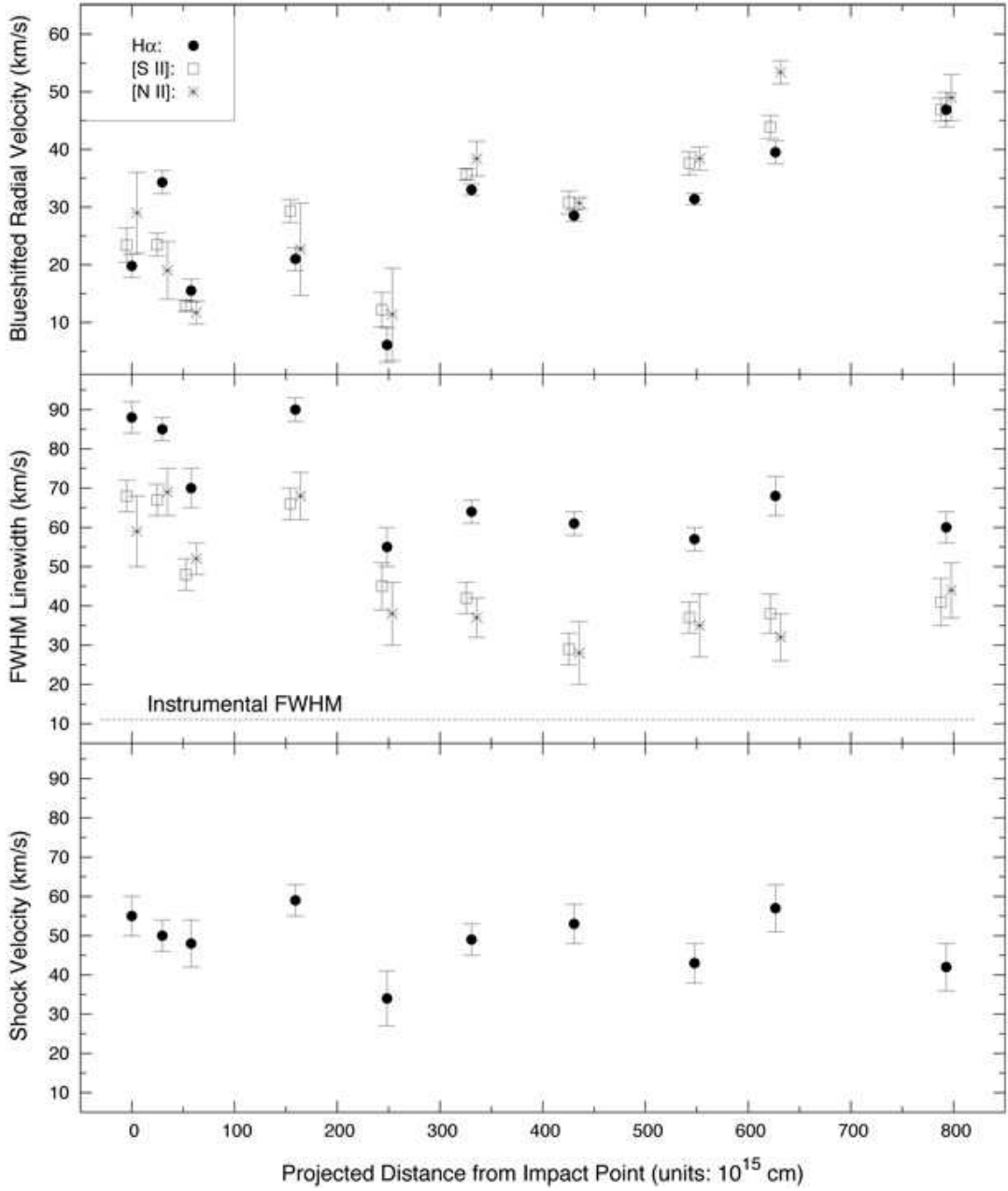}
\vskip 6.9in
\caption{
Kinematics of HH 110. Top: Blueshifted radial velocity of the regions along the slit
marked in Fig.~15, measured relative to the HH 110 molecular cloud.
The three emission lines shown for each region are offset slightly
in the x-direction for clarity. Middle: Observed FWHM of the emission line profiles
in Fig.~17. The H$\alpha$ linewidths are larger than those of the forbidden lines
owing to thermal broadening. Bottom: Shock velocities inferred for each region,
based on the thermal speed of H$\alpha$ in the postshock gas.
}
\end{figure}
\null\vfill\eject

\begin{figure} %nfig19
\def\thefigure{19}
\null\includegraphics{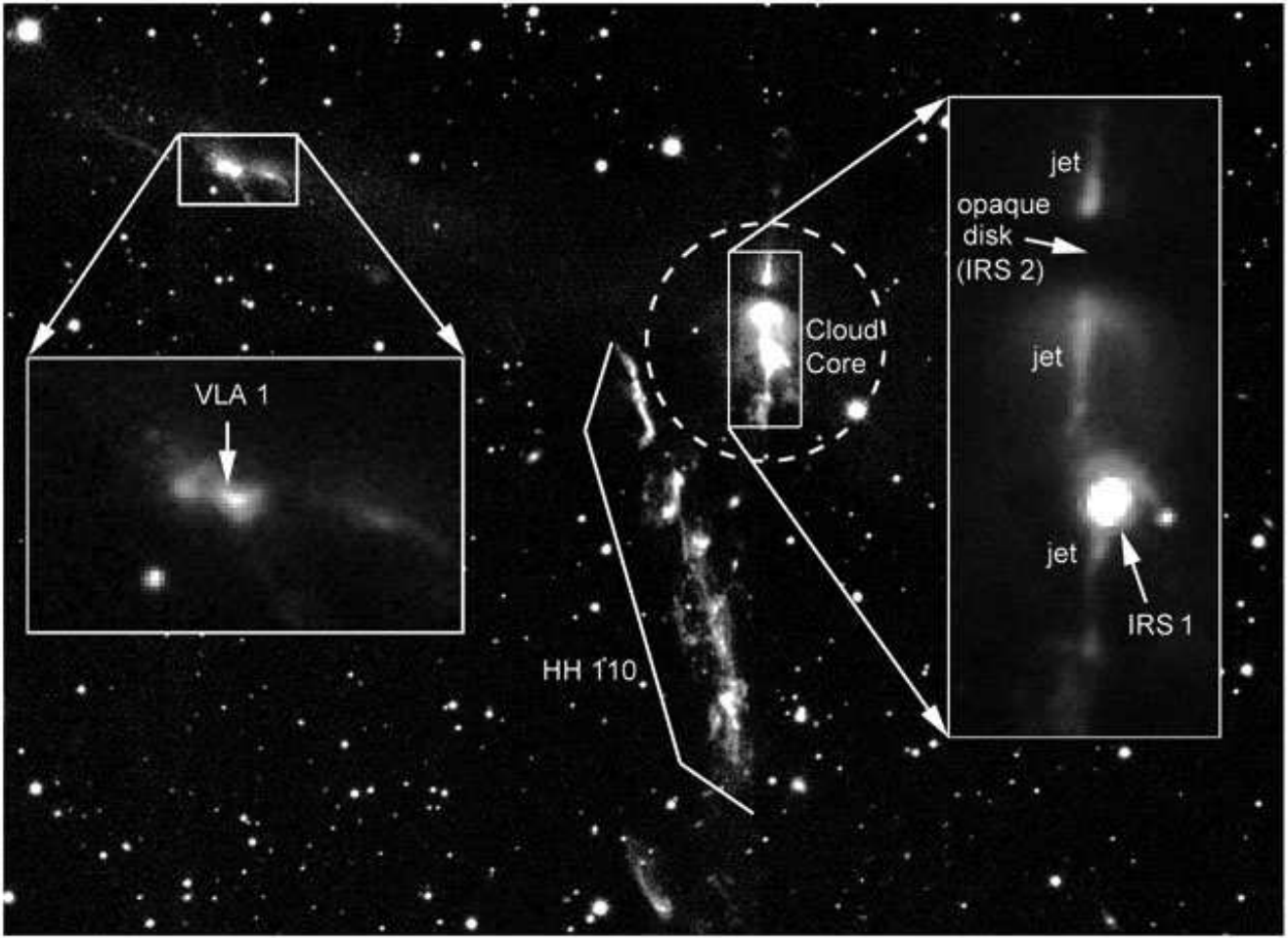}
\vskip 0.0in
\caption{The HH~110 region, imaged in the H$_2$ 2.12$\mu$m line with the NEWFIRM camera. 
The HH~110 flow originates at the VLA 1 source \citep{choi06} and radiates in H$_2$ after it
deflects from a molecular cloud core. Two sources, IRS~1 and IRS~2, drive nearly parallel jets within 
the core itself. IRS~2 appears to be an opaque flared disk seen nearly edge-on.
}
\end{figure}
\null\vfill\eject

\begin{figure} %nfig20
\def\thefigure{20}
\null\includegraphics{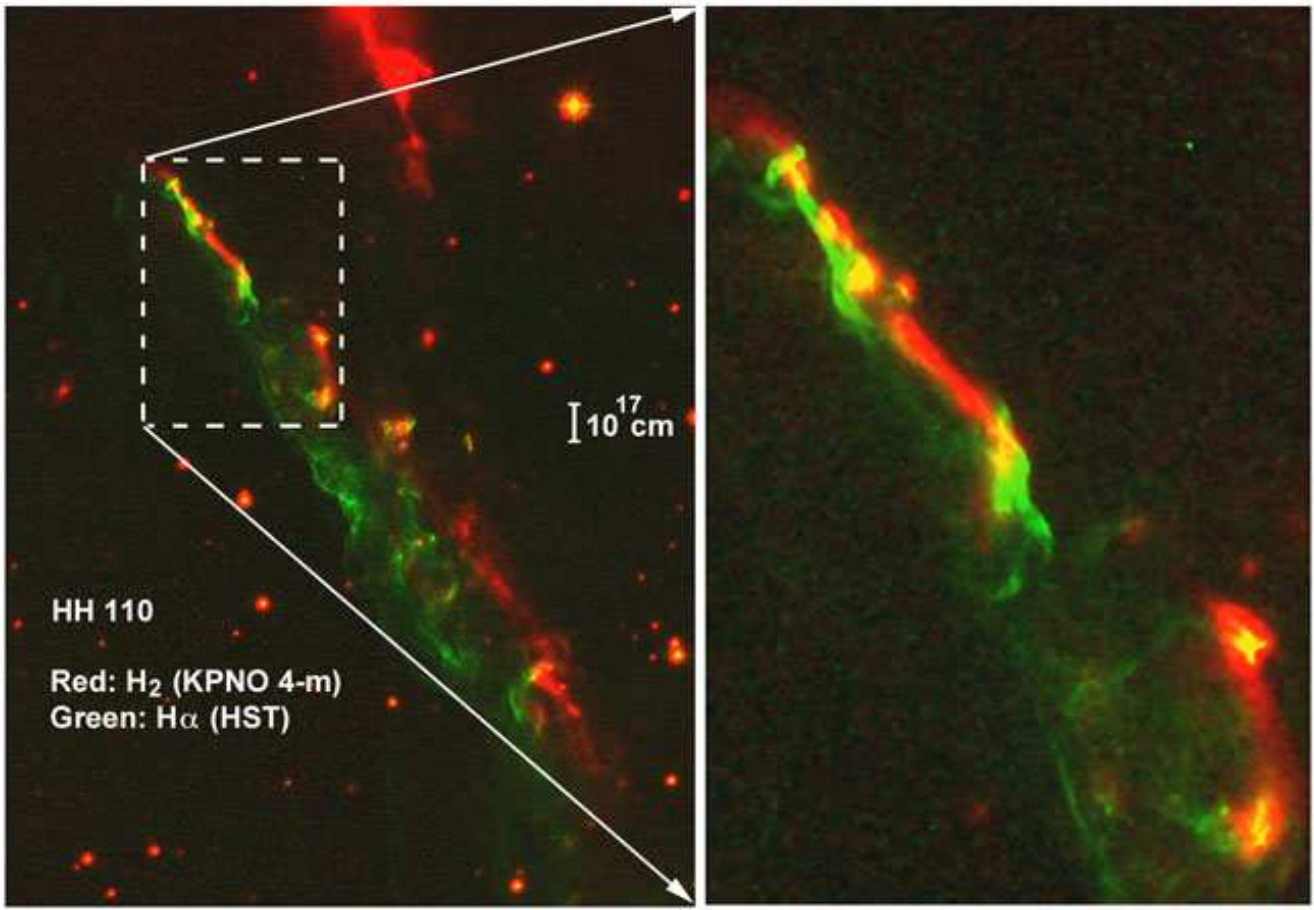}
\vskip 0.0in
\caption{Image of H$_2$ (red, from NEWFIRM), and H$\alpha$ (green, from the Hubble Space Telescope archive)
of HH~110. The H$_2$ emission is systematically offset to the right the H$\alpha$, consistent with
material being dragged from the molecular cloud core by the glancing collision with a jet.
}
\end{figure}
\null\vfill\eject

\begin{figure} %nfig21
\def\thefigure{21}
\null\includegraphics{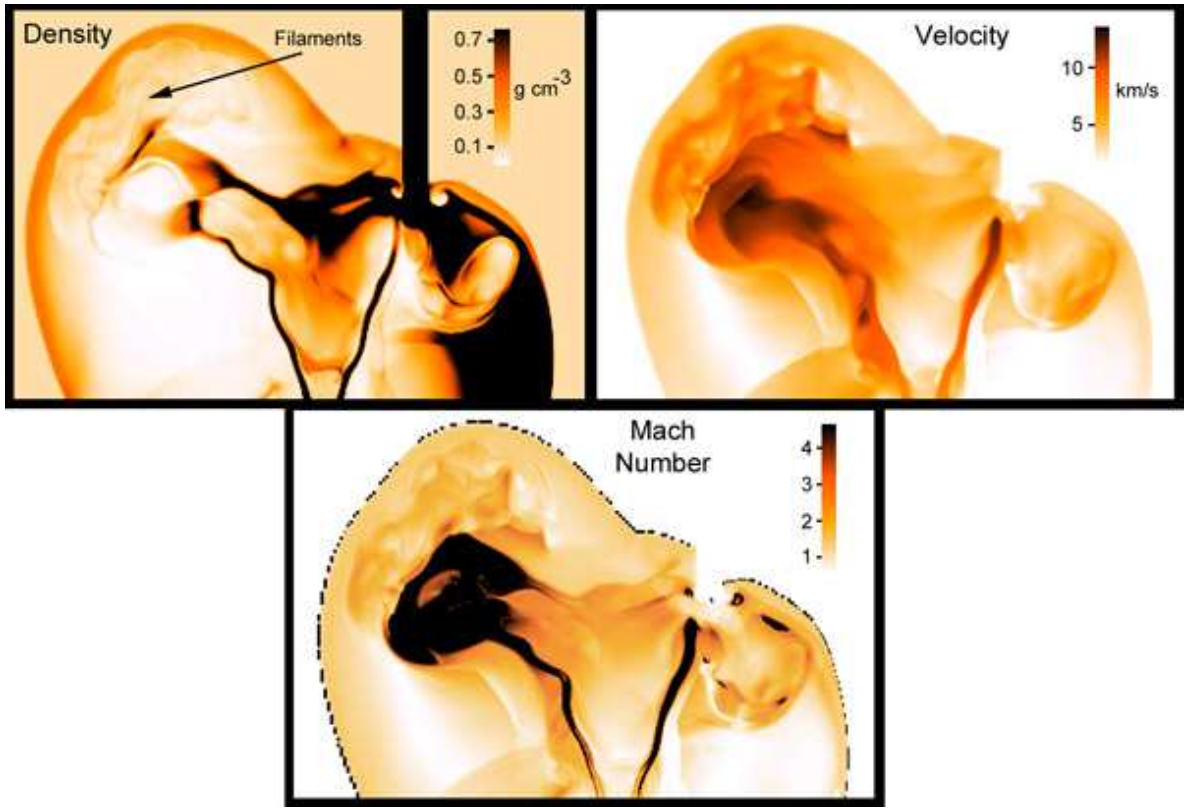}
\vskip 0.0in
\caption{RAGE simulations of the dynamics within the deflected experimental jet.
The figure shows the density, velocity, and local Mach number at each point within
the plane of symmetry of the experiment at 200~ns. The filaments within the working
surface of the deflected bow shock are discussed in the text.
}
\end{figure}
\null\vfill\eject

\begin{figure} %nfig22
\def\thefigure{22}
\null\includegraphics{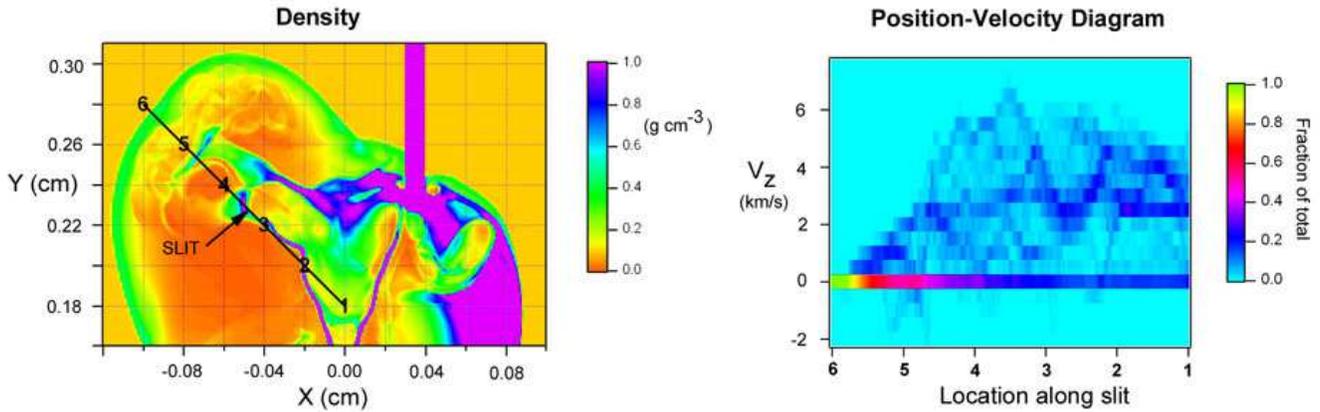}
\vskip 0.0in
\caption{Synthetic position-velocity diagram of the laser experiment. The density
image at left is a slice through the symmetry plane (z=0), which by definition contains the axis
of the jet and the center of the ball. Negative values of z are into the paper.
At each position (x$_\circ$,y$_\circ$) along the `slit', the position-velocity diagram at right 
gives a probability distribution for the radial velocity V$_Z$ of material located 
at (x$_\circ$, y$_\circ$, z$<$0). The large amount of material at V$_Z$=0 simply represents
undisturbed foam. Material at (x$_\circ$, y$_\circ$, z$>$0)
produces an identical p-v diagram (not shown) but with the sign of V$_Z$ reversed.}
\end{figure}

\null\vfill\eject

\normalsize

\end{document}